\documentclass[10pt,journal,compsoc]{IEEEtran}

\PassOptionsToPackage{dvipsnames,svgnames}{xcolor}
\usepackage{amsmath,amsfonts}
\usepackage{algorithmic}
\usepackage{array}
\usepackage{stfloats}
\usepackage{lipsum}                    
\usepackage{mathptmx}                  
\usepackage{mwe}                       
\usepackage{url}
\usepackage{verbatim}
\usepackage{fontawesome5}
\usepackage{colortbl}
\usepackage{booktabs}                  
\usepackage{mathptmx}                  
\usepackage{footnote}
\usepackage[hang,flushmargin]{footmisc}
\usepackage{listings}
\usepackage{multirow}
\usepackage{helvet}
\usepackage{soul}
\usepackage{color}
\usepackage[framemethod=TikZ]{mdframed}
\usepackage{fontawesome5}
\usepackage[export]{adjustbox}
\usepackage{svg}
\usepackage[caption=false,font=footnotesize,labelfont=sf,textfont=sf]{subfig}
\usepackage{cite}
\usepackage{makecell}

\definecolor{fulvous}{rgb}{0.2, 0.2, 0.90}

\newcommand{\addt}[1]{#1} 

\newenvironment{insight}[2][LightSteelBlue]{%
    \ifstrempty{#2}%
    {\mdfsetup{%
        frametitle={%
            \tikz[baseline=(current bounding box.east),outer sep=0pt]
            \node[anchor=east,rectangle,fill=#1!80,rounded corners]
            {\strut~Insight~};}}
    }%
    {\mdfsetup{%
        frametitle={%
            \tikz[baseline=(current bounding box.east),outer sep=0pt]
            \node[anchor=east,rectangle,fill=#1!80,rounded corners]
            {\strut~#2~};}}%
    }%
    \mdfsetup{innertopmargin=-4pt,linecolor=#1!80,backgroundcolor=#1!20,%
        linewidth=2pt,topline=true,roundcorner=5pt,%
        frametitleaboveskip=\dimexpr-\ht\strutbox\relax
    }
    \begin{mdframed}[]\relax}
    {\end{mdframed}}

\definecolor{new-green}{rgb}{0.104,0.667,0.229}
\newcommand{\add}[1]{{\color{new-green}\textbf{{#1}}}}

\def\BibTeX{{\rm B\kern-.05em{\sc i\kern-.025em b}\kern-.08em
    T\kern-.1667em\lower.7ex\hbox{E}\kern-.125emX}}
\begin{document}
\title{Drillboards: Adaptive Visualization Dashboards for Dynamic Personalization of Visualization Experiences}

\author{Sungbok Shin*, Inyoup Na*, and Niklas Elmqvist,~\IEEEmembership{Fellow,~IEEE}
\IEEEcompsocitemizethanks{
    \IEEEcompsocthanksitem Sungbok Shin is with Inria-Saclay and Universit\'e Paris-Saclay, Saclay, France. The work was done while the author was affiliated with the University of Maryland, College Park.
    E-mail: sungbok.shin@inria.fr 
    \IEEEcompsocthanksitem Inyoup Na is with the Department of Computer Science and Engineering at Korea University, Seoul, South Korea.
    E-mail: windy9898@korea.ac.kr
    \IEEEcompsocthanksitem Niklas Elmqvist is with the Department of Computer Science at Aarhus University, Aarhus, Denmark.
    E-mail: elm@cs.au.dk}
    \thanks{Manuscript received XXX XX, 2024; revised XXX XX, 2024.\\
    * denotes equal contributions.}
}

\markboth{IEEE Transactions on Visualization and Computer Graphics}{Shin \MakeLowercase{\textit{et al.}}: \techname}

\IEEEtitleabstractindextext{%
\begin{abstract}
We present drillboards, a technique for adaptive visualization dashboards consisting of a hierarchy of coordinated charts that the user can drill down to reach a desired level of detail depending on their expertise, interest, and desired effort. 
This functionality allows different users to personalize the same dashboard to their specific needs and expertise. 
The technique is based on a formal vocabulary of chart representations and rules for merging multiple charts of different types and data into single composite representations. 
The drillboard hierarchy is created by iteratively applying these rules starting from a baseline dashboard, with each consecutive operation yielding a new dashboard with fewer charts and progressively more abstract and simplified views. 
We also present an authoring tool for building drillboards and show how it can be applied to an agricultural dataset with hundreds of expert users. 
Our evaluation asked three domain experts to author drillboards for their own datasets, which we then showed to casual end-users with favorable outcomes.
\end{abstract}

}

\markboth{Transactions on Visualization and Computer Graphics}%
{How to Use the IEEEtran \LaTeX \ Templates}

\maketitle

\section{Introduction}
\label{sec:intro}

\begin{figure*}[tbh]
    \centering
    \includegraphics[width=\textwidth]{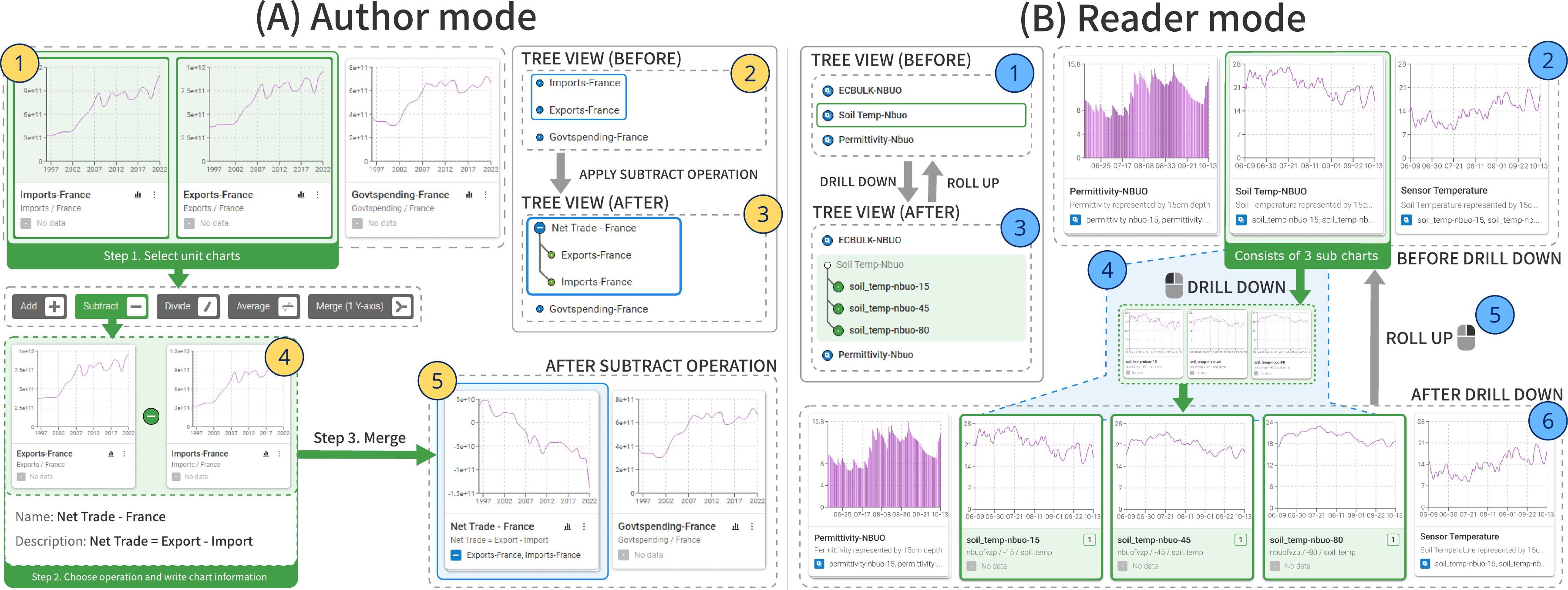}
    \caption{\textbf{Drillboards} are adaptive visualization dashboards consisting of a hierarchy of coordinated charts. 
    A drillboard has two modes: authoring (A) and reading (B). 
    The hierarchy is designed in authoring mode, and then explored in reading mode.
    The author-designed hierarchical structure is shown in the hierarchy browser. 
    In the reading mode, the user can drill down a chart to expand it into its constituent child charts.
    The user can also roll up the hierarchy to collapse a few charts.
    Here is how the author mode (A) works: \textcircled{1} The author needs to select charts of their interest. \textcircled{4} After selecting the operation to merge the charts and adding descriptions and title, \textcircled{5} the chart is merged. The treeview also is updated based on the changes made (\textcircled{2}, \textcircled{3}).
    Here is how the reader mode (B) works: \textcircled{1} The reader first sees a chart of their interest, and left-clicks the chart. Then, \textcircled{6} the original chart disappears, and 3 children charts appear instead, in a highlighted manner. \textcircled{5} By right-clicking any children charts, the charts roll-up to transform into the parent chart. Here too, the treeview (\textcircled{1}, \textcircled{3}) is updated based on the changes made in the chart view.  } 
    \label{fig:teaser}
\end{figure*}

\begin{figure*}[!tb]
    \centering
    \subfloat[\textbf{Author mode.} \textcircled{1} is the treeview which shows hierarchical relationships between charts, and the title of charts. \textcircled{2} is where users can select the charts of their interest. When multiple charts are selected from the operator implementations \textcircled{3}, authors can choose an operation, view the result, define a new title, and write explanations.
    Users can view the x- and y-labels by toggling on in the button in \textcircled{4}.
    ]{
        \frame{\includegraphics[height=4.07cm, keepaspectratio]{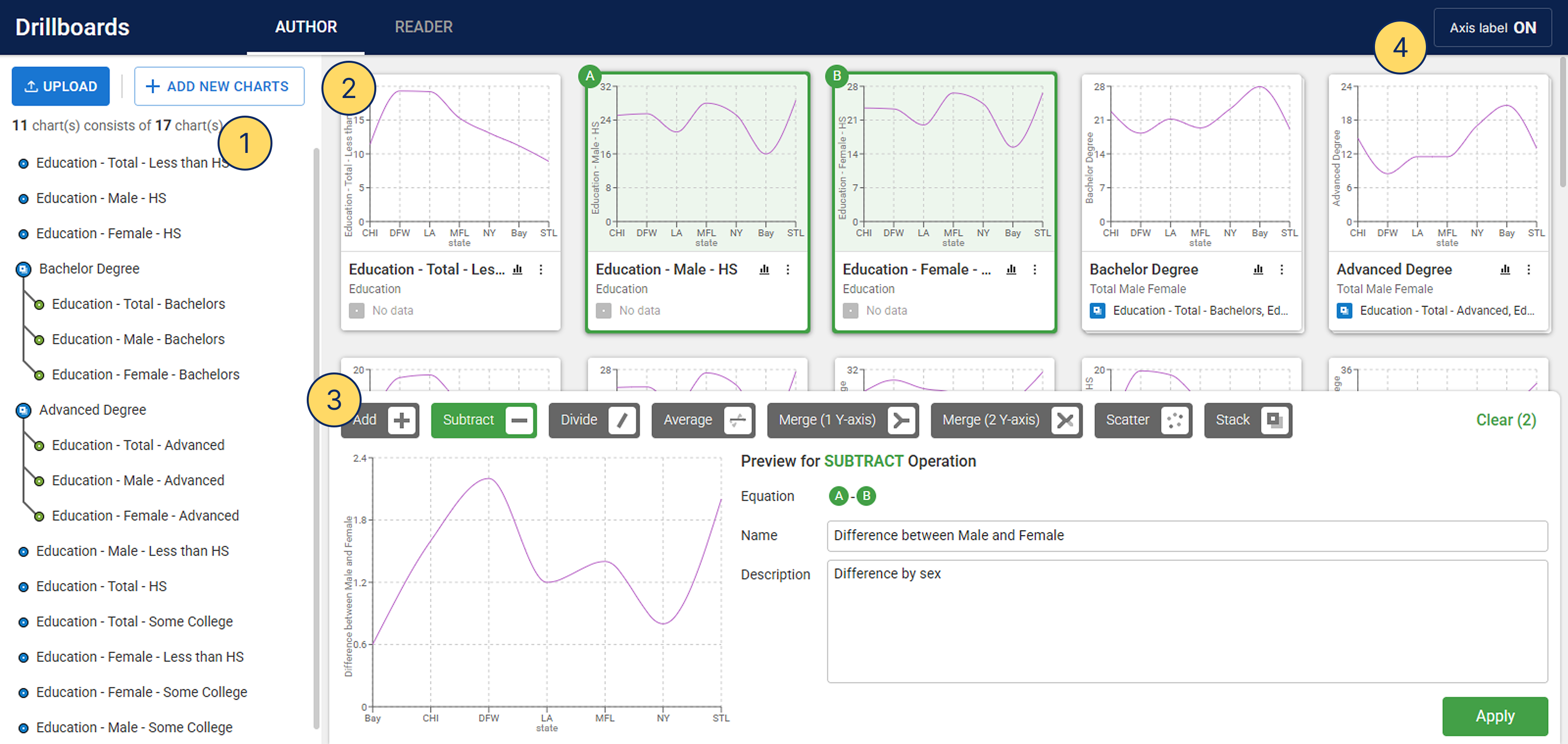}}
        \label{subfig:author-mode-overview}
    }
    \hspace{0.03in}
    \subfloat[\textbf{Reader mode.} From \textcircled{1}, readers can select drillboards of their interest. Drillboards offers different versions for readers with varying levels of domain knowledge (see \textit{defining views} in Sec.~\ref{subsec:authoring-mode}). \textcircled{2} is a series of highlighted unit charts drilled down from a parent chart. These parent-child relationship can be found in the treeview in \textcircled{3}. ]{
        \frame{\includegraphics[height=4.07cm, keepaspectratio]{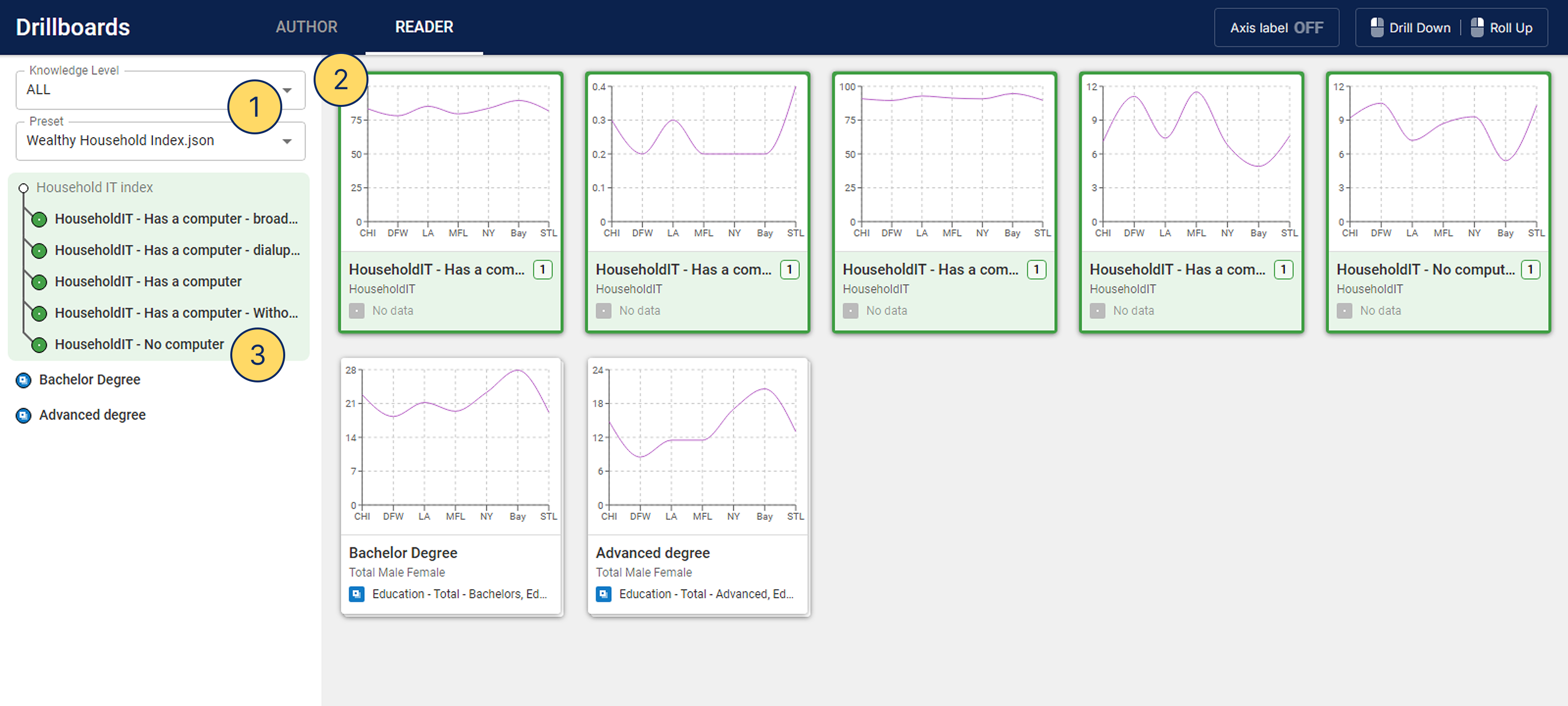}}
        \label{subfig:reader-mode-overview}
    }
    \caption{\textbf{Author and reader modes in DrillVis.} In the author mode, authors can create aggregate hierarchies using unit charts. In reader mode, users can drill down into visualizations to explore the details hidden beneath the selected chart.}
    \label{fig:author-readermodes}
\end{figure*}

\IEEEPARstart{V}{isualization} dashboards are becoming increasingly popular for communicating data to the general public as well as supporting decision-making in virtually every knowledge industry. 
A recent study~\cite{sarikaya19dashboards_tvcg} surveyed 83 existing dashboards ``in the wild'' and qualitatively categorized this corpus based on purpose, audience, visual features, and semantics. 
Central to this study is the idea that each and every dashboard is designed for a specific audience and purpose, making them essentially immutable in nature even if the data being visualized is dynamic and streaming. 
But what if the user wants to do something different with the dashboard than it was intended for? 
What if the dashboard is intended for a different audience than the current user?

We propose \textit{drillboards}: an adaptive user interface (AUI)~\cite{Greenberg1985, AUIbook1993} technique for dynamic visualization dashboards that can be \textit{drilled into} to accommodate different purposes, audiences, and efforts for each specific user. 
Instead of a static and immutable dashboard structure consisting of a grid of coordinated charts, a drillboard is essentially a hierarchy of charts. 
At the bottom of this hierarchy is the baseline dashboard, which consists of all of the different charts displaying the data at its highest detail. 
Each consecutive hierarchy level above the baseline merges two or more charts, yielding a less detailed representation. 
The root of the drillboard hierarchy is thus a single chart that represents the entire dashboard in a compact and easily glanceable representation. 
Drilling down into the hierarchy allows a user to adapt their view of the dashboard depending on their expertise, purpose, and desired effort. 
A drillboard can also provide predefined aggregation levels, such as ``novice,'' ``intermediate,'' and ``expert.''

While the hierarchy can be generated in an ad hoc manner, imposing structure on the process makes the drillboard aggregation algorithm easier to navigate and use.
To this end, we present a formal vocabulary of dashboard chart representations as well as rules for merging two or more charts into one chart (either using the same visual representation or a different one). 
A drillboard can thus be created by applying these rules iteratively starting from the baseline dashboard.
Of course, authoring a drillboard is significantly more complex than authoring a normal visualization dashboard with a static structure. 
To support this task, we present an authoring environment---\textsc{DrillVis}---for creating web-based drillboards to visualize multidimensional data. 
DrillVis lets the user load a multidimensional dataset and then create a normal static visualization dashboard (the baseline). 
Then it provides a series of choices for how to construct the drillboard hierarchy by applying the available rules on the specific dashboard. 
The decision for which rule to apply is left to the user. 
The user can also specify predefined labels in the hierarchy, allowing the end-user to easily navigate to specific views of the drillboard. 
Finally, the drillboard can be exported into a read-only form for distribution to end-users, who can both navigate the hierarchy manually or use the predefined labels.

We demonstrate the utility of the DrillVis authoring environment by allowing 3 domain experts who regularly engage and interact with data to author drillboards for novices. 
Following that, we recruit 10 casual end-users to assess the drillboards created by these experts.
Our findings reveal that drillboards serve effectively as a communication tool, especially for domain experts aiming to convey the context and provenance of the data they visualize. 
Furthermore, the results show that novices can swiftly grasp the experts' intentions through this platform.
\medskip

\textbf{Contributions.} We claim the following contributions:

\begin{itemize}
\item Our drillboards algorithm, which allows an effective exploration of visualization via drill-down and roll-up of aggregated visualizations for different levels of expertise;
\item The \textbf{DrillVis} system that allows authoring of aggregate drillboards (author mode) and exploration of data via drill-down and roll-up interactions (reader mode); and
\item Favorable results from a qualitative user study involving 3 domain experts and 10 end-users assessing the system. 
\end{itemize}

\section{Background}
\label{sec:background}

We describe the related work from four perspectives: (1) visualization for multiple views, (2) dashboards in visualization literature, (3) visualization of hierarchical aggregation, and (4) authoring dashboards and multiple views.
At the end of each subsection we specify our contributions in relation to prior art.

\subsection{Multiple Views}

The academic community has contributed to dashboards by studying the concept of multiple charts.
Baldonado et al.~\cite{DBLP:conf/avi/BaldonadoWK00} present guidelines for effectively integrating multiple perspectives of data to support comprehensive analysis.
Roberts~\cite{Roberts2007} report on the state of the art in the field of \textit{coordinated multiple views}, a key aspect of visualization dashboard design.
Javed and Elmqvist~\cite{Javed2012a} present a theoretical framework for dynamically composing visualizations to support flexible exploration of multidimensional data.
Finally, Qu and Hullman~\cite{DBLP:journals/tvcg/QuH18} demonstrate how these ideas apply to dashboards by examining the role of consistency in multi-view visualizations for effective visualization composition.

We draw on many of these ideas for our drillboards technique, but do not claim any contribution to multiple views.

\subsection{Dashboards in Visualization Literature}
\label{subsec:back-interactivedash}

Dashboards have emerged as a popular visualization format for exploring and monitoring data, facilitating easy overview of multiple charts for non-expert users.
This effectiveness has led to the proliferation of web-based visual analytics tools across diverse domains, including business (e.g.,~\cite{noonpakdee18dashboardbi, conf/chi/BadamZSEE16,samrose21meetingcoach_chi, fu18tcal_chi, reuter15xhelp_chi}), urban analytics (e.g.,~\cite{hohman19gamut, jin23trafficvis_tvcg, garciazanabria22cripav_tvcg, yang23epimob_tvcg,cao18voila}), medicine (e.g.,~\cite{zhang21mapcovid, DBLP:journals/tvcg/ElshehalyRBMAGR21, choi22mitovis_tvcg, li20maravis_chi, vankollenburg18healthprof_chi}), learning (e.g.,~\cite{alzoubi21educdash, mcneill23zombievis_chi, thanyadit23tutorvis_chi, mendez21academic_chi}), explainable AI (XAI) (e.g.,~\cite{hoque23visualconcept_tvcg, elhamdadi22affectivetda_tvcg, huang23conceptexplainer_tvcg, wang22vaod_tvcg,yan20silva_chi}), and more. 

In stark contrast to their widespread popularity, academic interest in understanding their usage and design patterns has been relatively limited~\cite{sarikaya19dashboards_tvcg}, and it is only in recent years that the visualization community has begun to delve into this underexplored area.
Sarikaya et al.~\cite{sarikaya19dashboards_tvcg} provide a comprehensive characterization of dashboards based on their designs, purposes, audience, and data semantics, drawing from a thorough survey of 83 dashboards.
Building on this foundation, Bach et al.~\cite{bach23dashboardpatterns_tvcg} conducted a detailed examination of 144 examples to yield common dashboard design patterns.
From these studies, they propose a simplified model for understanding the tradeoffs involved in dashboard design, focusing on four key features: screen space, number of pages, abstraction, and interaction.
Their findings suggest that achieving optimization in one aspect often requires sacrifices in others.

Drillboards offer a novel approach to enhance the effectiveness of dashboards as communication tools by addressing two critical aspects:
(1) the imperative for end-user flexibility, as emphasized in previous research~\cite{sarikaya19dashboards_tvcg}, and
(2) the challenge of mitigating information loss, as highlighted by recent studies~\cite{bach23dashboardpatterns_tvcg}, without resorting to increased screen space or additional pages.
Drillboards facilitate end-user flexibility through a dynamic drill-down/roll-up interface, empowering users to explore data comprehensively. 
We claim that adaptable drill-down/roll-up can effectively counteract information loss, particularly for single-page dashboards, thereby enhancing data presentation and comprehension.

\subsection{Visualizing Hierarchical Aggregation} 
\label{subsec:data-aggregation}

A wealth of research has been dedicated to effectively visualizing and summarizing data into hierarchical aggregates, driven primarily by the need for managing large-scale datasets efficiently.
This pursuit began with the advent of data cubes~\cite{Stolte2003, lins13nanocubes, pahins17hashedcubes, wang21neuralcubes}, which introduced algorithms aimed at optimizing computational resource utilization---notably memory and CPU usage---when handling extensive datasets.
Additionally, effective summarization techniques for large-scale datasets have been explored~\cite{doraiswamy18taxivis, DBLP:journals/tvcg/McDonnelE09}.
Another focus is on presenting data structure and relationships effectively, exemplified by methods such as trees and treemaps~\cite{shneiderman92treemap, johnson91treemaps, balzer05voronoitreemap, bederson02oqtreemaps}.
Similarly, thematic representations, such as streamgraph-based visualizations~\cite{DBLP:journals/tvcg/ByronW08, DBLP:journals/tvcg/GadJGEEHR15, havre02themeriver}, have been employed to illustrate topic evolutions~\cite{wei10tiara, chi11textflow, heimerl16citerivers}.

Another approach to aggregating data involves customizing its structure through hierarchical aggregation.
Elmqvist and Fekete~\cite{Elmqvist2010} offer guidelines for this method and introduce various interaction techniques that facilitate navigation within the structured data.
A common interaction technique used for visualizing this relationship is zooming and panning, frequently employed in systems aiming to adjust window view position and resolution~\cite{choi18topicontiles, kim17topiclens}.
Another prevalent interface allows data exploration through drill-down and roll-up~\cite{piringer10hierscatterplot, yang02interring,henry07nodetrix, marcus11twitinfo, gotz20vahieragg}; expanding or collapsing aggregate data items, respectively, to increase or decrease the amount of data shown.

While the hierarchical presentation of data in dashboards has been explored previously~\cite{bach23dashboardpatterns_tvcg}, the incorporation of drilling down and rolling up components charts within dashboards represents a novel approach.
We argue that integrating hierarchical aggregation into dashboards offers a dual advantage: it optimizes space utilization effectively while also serving as a powerful tool for delivering personalized visualization experiences.

\subsection{Authoring Dashboards and Multiple Views}
\label{subsec:personalization}

Because of their prevalence in business intelligence and commercial data analysis, dashboard design has long been a focus in the business community~\cite{few2006information}.
As a result, we identify tools that support advanced authoring functionality.
Tableau (née Polaris~\cite{stolte02polaris_tvcg}) offers a powerful dashboard authoring interface integrated with its visualization and analytics capabilities.
Microsoft Power BI\footnote{\url{https://www.microsoft.com/en-us/power-platform/products/power-bi}} is another popular choice for dashboard authoring that integrates seamlessly with other Microsoft tools and offers robust data analytics and visualization features.
Launched in 2016, Google Data Studio was a free tool for creating interactive dashboards using data from various sources, including Google Analytics, Google Ads, etc.
In 2022, the software suite was rebranded as Looker Studio\footnote{\url{https://lookerstudio.google.com/}} after Google's successful \$2.6B acquisition of data analytics company Looker in 2019.
Finally, the Qlik Sense\footnote{\url{https://www.qlik.com/us/products/qlik-sense}} data analytics platform enables users to create interactive dashboards and reports from visualizations created during data analysis.

Several dashboard authoring mechanisms have been proposed by the academic community.
Visualization mosaics~\cite{DBLP:journals/cgf/MacNeilE13} enables composing multidimensional dataset visualizations using basic components such as barcharts and scatterplots organized in a space-filling slice-and-dice mosaic layout with the ability to dynamically change visual representations.
Keshif~\cite{Yalcin2017b} allows for creating interactive visualization dashboards by supporting exploratory data analysis and providing many visual encoding options, dynamic interactivity, and collaborative features.
The framework has an authoring mode for constructing the dashboard and a viewing mode for distribution to end-users.

The most recent work in this space leverages automated methods and AI to aid dashboard design. 
StoryFacets~\cite{DBLP:journals/ivs/ParkSZDRE22} use automatically generated dashboards as one of three storytelling mechanisms to report on data analysis outcomes.
LADV~\cite{DBLP:journals/tvcg/MaMGHZXDWC21} offers a rapid conceptualization approach for constructing dashboard templates from exemplars using a deep learning-based model, enhancing efficiency in dashboard visualization authoring for data-intensive applications.
Similarly, MultiVision~\cite{DBLP:journals/tvcg/WuWZHZQZ22} provides a deep learning-based method that automates the design of analytical dashboards by recommending meaningful combinations of data columns and multiple charts to support data selection and composition.
MEDLEY~\cite{DBLP:journals/tvcg/PandeySS23} automates dashboard composition by recommending collections of visualizations and filtering widgets based on specific analytical intents, simplifying dashboard authoring process.
BOLT~\cite{Srinivasan2023} automates dashboard authoring through natural language, enabling users to express their objectives and receive relevant recommendations to streamline dashboard creation.

Most relevant to our drillboards work is QualDash~\cite{DBLP:journals/tvcg/ElshehalyRBMAGR21}, which automates and adapts dashboard design for healthcare quality improvement by dynamically configuring visualizations based on a metric specification structure, enhancing ease-of-use and usefulness in diverse healthcare contexts.
Their dashboard generation engine can dynamically generate a new dashboard from the same data source based on customizable visualization ``cards.''





\addt{Drillboards provides a unique approach to hierarchically grouping charts into aggregates to yield an adaptive user interface.}
While automated methods may be utilized to customize a dashboard for a particular user, such as in QualDash~\cite{DBLP:journals/tvcg/ElshehalyRBMAGR21}, a drillboard at its core remains invariant---it just offers more or less detail depending on the needs of the user.
This promotes consistency across users and expertise levels, and also facilitates seamless \textit{upskilling} where users grow in expertise and accordingly drill deeper for more detail (as well as downskilling, in case of a user not having visited a dashboard for a longer period of time).


\section{Design Framework: Drill-down Dashboards}
\label{sec:drillvis}

\textit{Drill-down dashboards}, or \textit{drillboards}, are an adaptive user interface~\cite{Greenberg1985, AUIbook1993} for dynamic visualization dashboards that provide multiple levels of detail through an aggregation hierarchy that users can drill-down and roll-up~\cite{Elmqvist2010}.
Here we design the design goals, rationale, and formal operations of the approach.

\subsection{Design Rationale}
\label{sec:rationales}

Visualization dashboards are forms of communication-minded visualizations~\cite{Viegas2006}; they are designed for use by multiple users~\cite{sarikaya19dashboards_tvcg}.
Consider a typical visualization dashboard created for use in an organization or published on a website.
While the former typically has a more controlled user audience than the latter, it still remains that the multiple users that frequent a visualization will have varying expertise. 
This leads us to our first design goal (DG1):

\begin{insight}{DG1: One Dashboard---\addt{Adaptive to Users}}
    \addt{We want to provide the same visualization dashboard content (though, with different expertise levels) for users with varying domain and visualization expertise.}
\end{insight}

Second, not only will multiple users need to use the same dashboard, but they may have different tasks that they want to complete at different times when using the dashboard. 
Some tasks deal with quick overviews, whereas others are more deliberate and focused. 
Put differently:

\begin{insight}{DG2: One Dashboard---Multiple Tasks}
    The task to complete using the same dashboard varies, including the level of detail, time spent, and required focus.
\end{insight}

Third, the point of an adaptive user interface---beyond adapting to the user's context, task, and needs---is to support users upskilling \addt{(or downskilling---when the user did not use the tool for a long time, and needs time to familiarize with the tool)} using the same interface.
An ideal visualization dashboard would support users learning its use over time:

\begin{insight}{DG3: One Dashboard---Changing Skill}
    Users may increase or decrease in skill over time, yet should be able to use the same visualization dashboard.
\end{insight}

Finally, authoring such a one-size-fits-all adaptive dashboard must be relatively straightforward. 
\addt{For this, the tool must be able to effectively guide users to author, and provide a good overview of how the hierarchy is made}: 

\begin{insight}{DG4: One Dashboard---Easy Authoring}
    Adaptive dashboards that support many users, \addt{charts} and tasks as well as changing skill must still be easy to author.
\end{insight}

\begin{figure*}[htb]
    \centering
    \subfloat[Labeling~\faFlag]{
        \includegraphics[height=3.5cm]{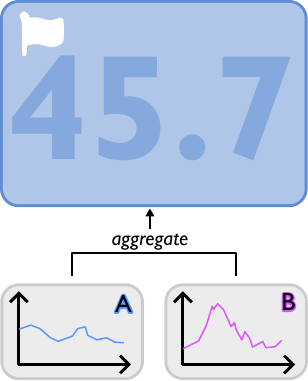}
        \label{fig:agg-label}
    }
    \subfloat[Summarization~\faCalculator]{
        \includegraphics[height=3.5cm]{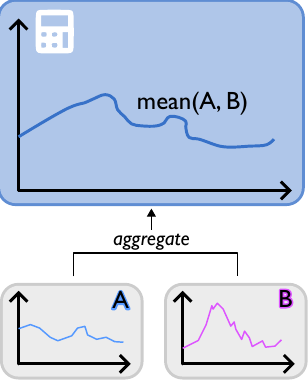}
        \label{fig:agg-summarize}
    }
    \subfloat[Archetype~\faStar]{
        \includegraphics[height=3.5cm]{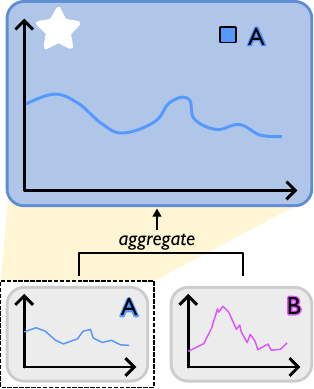}
        \label{fig:agg-archetype}
    }
    \subfloat[Projection~\faAngleDoubleDown]{
        \includegraphics[height=3.5cm]{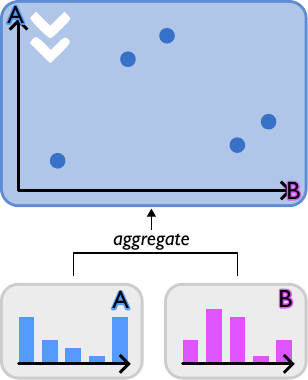}
        \label{fig:agg-project}
    }
    \subfloat[Juxtaposition~\faPlusCircle]{
        \includegraphics[height=3.5cm]{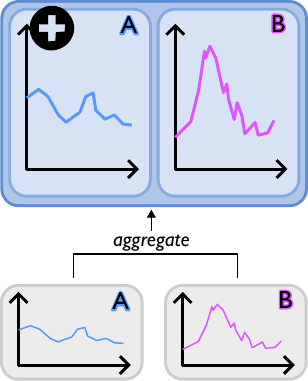}
        \label{fig:agg-juxtapose}
    }
    \subfloat[Overlay~\faLayerGroup]{
        \includegraphics[height=3.5cm]{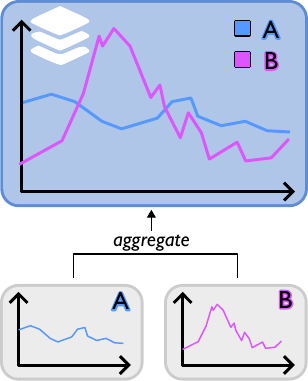}
        \label{fig:agg-overlay}
    }
    \caption{
    \textbf{Abstract aggregation operations.}
    Schematic illustrations of the abstract aggregate operations that can be applied to charts in a drillboard to build the aggregation hierarchy:
    \protect\subref{fig:agg-label}~labeling~\faFlag~the aggregate with text to represent the children; 
    \protect\subref{fig:agg-summarize}~summarizing~\faCalculator~the children with a data abstraction (e.g., mean); 
    \protect\subref{fig:agg-archetype}~selecting one child as archetype~\faStar~to represent them all;
    \protect\subref{fig:agg-project}~projecting~\faAngleDoubleDown~child data onto axes to form a scatterplot or parallel coordinate plot;
    \protect\subref{fig:agg-juxtapose}~juxtaposing~\faPlusCircle~multiple charts in the aggregate; and
    \protect\subref{fig:agg-overlay}~overlaying~\faLayerGroup~multiple data series in the same chart.}
    \label{fig:aggregation-operations}
\end{figure*}

\subsection{Model}

Let us consider a visualization dashboard as a dataset, a static collection of charts (\textit{chart atoms}~\faAtom) visualizing the data, and the chart layouts (positions and size).
Then let us define a \textit{drillboard} as a visualization dashboard that also contains an \textit{aggregation hierarchy} represented by a single \textit{pile}~\faObjectGroup~serving as its root.
A pile~\faObjectGroup~ is a recursive data structure that may either contain piles or chart atoms.
The visual representation of a pile is selected to represent its children.

The current drillboard \textit{view} is a sequence of piles and atoms representing the current aggregation state of the drillboard, and represent the charts that are currently visible on screen.
End-users can change the view by \textit{drilling-down} into a pile---which entails replacing the pile with its children in the view---or \textit{rolling-up} a chart atom or group---which entails replacing the chart and its siblings with their common ancestor.
Many drillboards also come with \textit{pre-defined views}; these are views that the drillboard author have created in advance, for example, to capture a specific level of detail or skill level.

\begin{figure}[tbh]
    \centering
    \subfloat[]{
        \includegraphics[width=0.45\linewidth,valign=t]{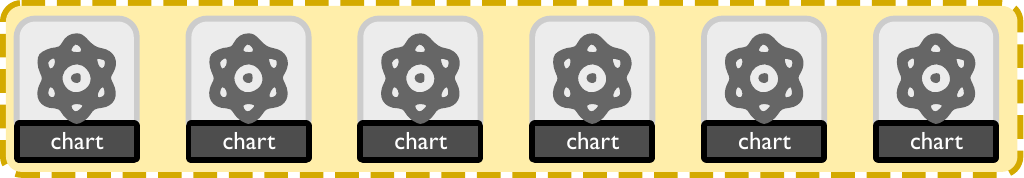}
        \vphantom{\includegraphics[width=0.45\linewidth,valign=t]{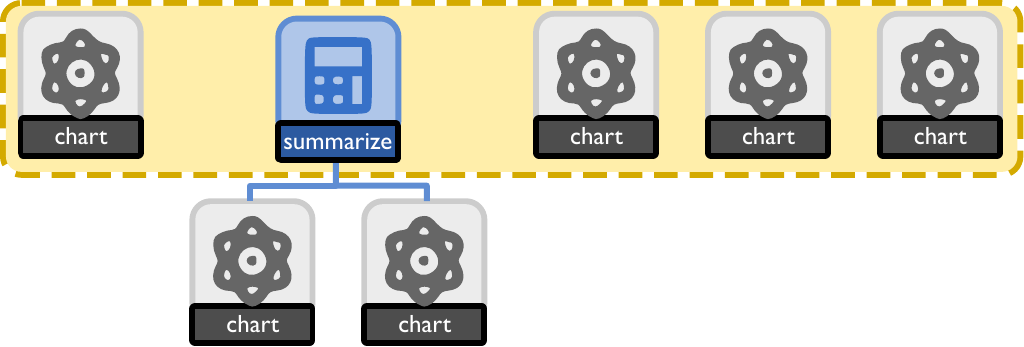}}
        \label{fig:agg-1}
    }
    \subfloat[Summarizing~\faCalculator~two charts.]{
        \includegraphics[width=0.45\linewidth,valign=t]{figures/agg-2.pdf}
        \label{fig:agg-2}
    }\\
    \subfloat[Archetyping~\faStar~two charts.]{
        \includegraphics[width=0.45\linewidth,valign=t]{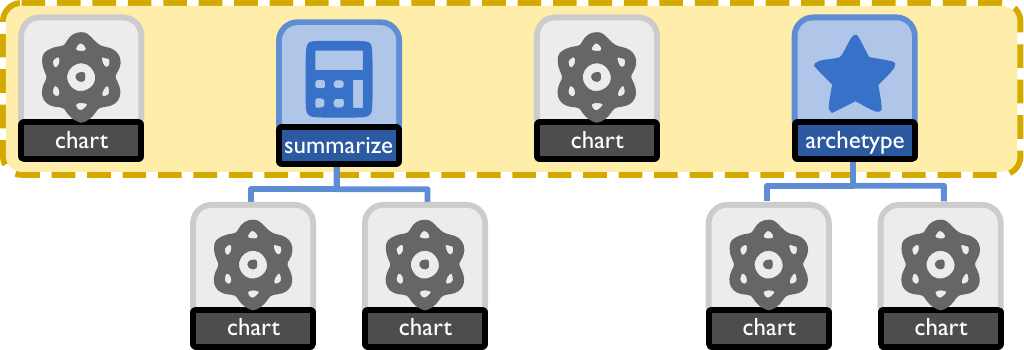}
        \vphantom{\includegraphics[width=0.45\linewidth,valign=t]{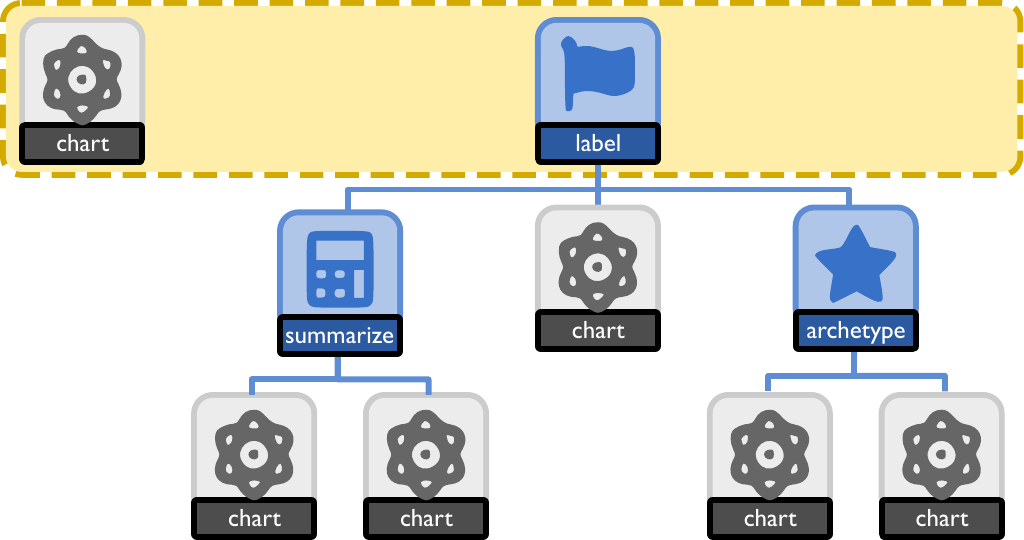}}
        \label{fig:agg-3}
    }
    \subfloat[Labeling~\faFlag~three charts.]{
        \includegraphics[width=0.45\linewidth,valign=t]{figures/agg-4.pdf}
        \label{fig:agg-4}
    }\\
    \subfloat[Archetyping~\faStar~two charts into the final aggregation hierarchy.]{
        \includegraphics[width=0.6\linewidth]{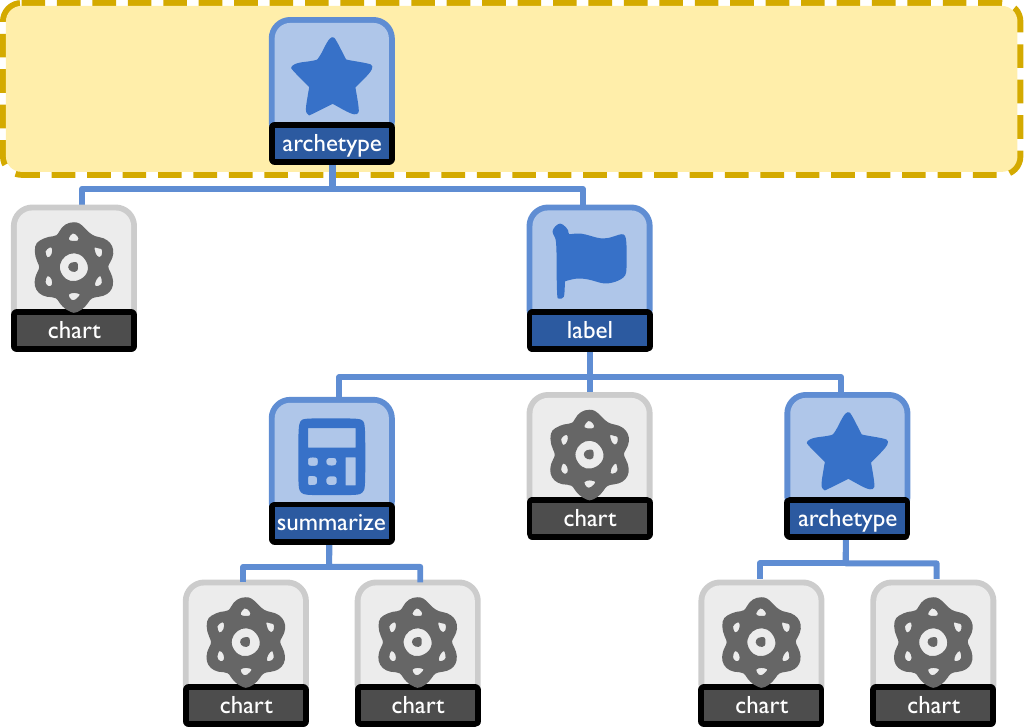}
        \label{fig:agg-5}
    }
    \caption{\textbf{Hierarchical aggregation.}
    Example of how an aggregation hierarchy is formed through a sequence of aggregation operations until a single pile remains as the root. 
    }
    \label{fig:aggregation}
\end{figure}

Authoring a drillboard entails starting with the chart sequence for a regular dashboard and then building the aggregation hierarchy by iteratively applying aggregations that replace two or more charts (atoms or groups) with a single pile.
Figure~\ref{fig:aggregation} shows an example.
Aggregation operations are described below.

The root of the aggregation hierarchy (the single node remaining in the yellow field in Figure~\ref{fig:agg-5}) and the original visualization dashboard (i.e., all of the chart atoms in the dashboard, see Figure~\ref{fig:agg-1}) are automatically pre-defined views.
The former could theoretically be used for a complete novice---just a single chart representing all the data in the dashboard---and the latter for an expert---all data visible and no information loss.
In addition to building the aggregation hierarchy, authors may also define current and pre-defined views \addt{(DG1)}.
If there is no current view, then the root of the aggregation hierarchy will form the initial view.
Additional pre-defined views can also be associated with the drillboard with a specific label.

\subsection{Aggregation Operations}
\label{subsec:aggregation_operations}

All aggregation operations are essentially \textsc{Merge} operations that replace one or more child charts---that could either be piles or chart atoms---with a single pile containing the selected children.
In aggregating charts, the new parent pile must be assigned a visual representation.
The choice of group representation is typically left in the hand of the drillboard designer (DG2).
There is typically---but not always---information loss inherent in this kind of merge.
Here we outline 6 possible merge operations (Figure~\ref{fig:aggregation-operations}.)
\addt{The list provided here consists of elements that support the contextualization of charts by integrating child charts. 
We mainly focus on three types of merge operations (DG2): (1) operations for structural organization and abstraction (SOA), (2) comparative and relational analysis (CRA), and (3) integrated representation (IR).
This list is not exhaustive and can be extended:}

\begin{itemize}
    \item[\faFlag]\textbf{Label (SOA):} Replace all of the chart children with a single label or text that summarizes their contents (i.e., \addt{not a form of data visualization}).
    This text could be a representative scalar, such as the mean, median, minimum, or maximum number in the underlying data.
    For example, charts showing the temperature in different regions could be replaced by the average temperature for the country.

    \item[\faCalculator]\textbf{Summarization (SOA):} Charts representing data of similar type can be summarized using data abstraction, such as calculating an average, sum, or difference between series, yielding a single new series. 
    
    \item[\faStar]\textbf{Archetype (SOA):} Select one of the chart children to serve as a representative for them all. 
    There is significant information loss in this kind of merge, but at least one of the original chart remains.

    \item[\faAngleDoubleDown]\textbf{Projection (CRA, IR):} Two or more data dimensions from child charts can be projected on the axis of a scatterplot (if two) or parallel coordinate plot (if more than two).
    Data can be mapped to a color scale---nominal or categorical to a categorical one, and an ordinal or quantitative one to a discrete or continuous ordered scale.
    
    \item[\faPlusCircle]\textbf{Juxtaposition (CRA):} Combine the child charts inside the extents of the parent pile as small multiples~\cite{DBLP:conf/apvis/JavedE12}.
    This preserves all of the information in the original charts, but may lead to increased visual clutter and reduced data resolution due to reduced space.

    \item[\faLayerGroup]\textbf{Overlay (IR):} Overlay the child charts on top of each other~\cite{DBLP:conf/apvis/JavedE12}.
    This operation requires that the child charts are of the same type; e.g., two line charts with a common vertical and horizontal axes.

\end{itemize}

\subsection{Addressing the Design Goals}

Drillboards consist of a variable number of charts: for example, one possible scenario is where there are a few, abstracted overview charts for novice users and many detailed charts for experts (\textbf{DG1}). This assumes that experts would favor to see more information, whereas novices, relatively less knowledgeable about the topic, would favor to view selectively guided information.
The aggregation hierarchy means that the number of charts to show can be controlled at fine granularity, allowing the same drillboard to be used for many different tasks (\textbf{DG2}).
The drill-down (and roll-up) operation enables a user to retrieve more (or less) detail as they learn (or forget) about the drillboard (\textbf{DG3}).
Finally, the aggregation hierarchy can be constructed through a sequence of aggregation operations in a dedicated authoring interface \textbf{(DG4)}; in fact, the hierarchy could even be built automatically using chart distance and hierarchical clustering.

\section{System: Drillboards}
\label{sec:drillboards}

We implement the drillboards technique in a web-based interactive visualization tool called \textsc{DrillVis}.
The tool has a basic visual interface consisting of drillboard view showing the current representation of the drillboard, and a tree view for seeing and navigating the aggregation hierarchy. 
The tool has two main interaction modes: (1) the author mode (see Figure~\ref{subfig:author-mode-overview}), and (2) the reader mode (Figure~\ref{subfig:reader-mode-overview}). 
Here we describe the interface, the aggregation operators, and the author and reader modes.

\begin{figure}[htb]
    \centering
    \subfloat[Card for \faAtom.]{
        \includegraphics[width=0.47\linewidth]{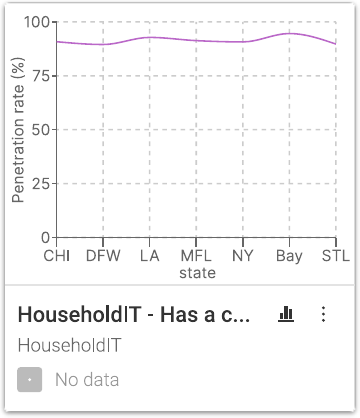}    
        \label{fig:chart-atom}
    }
    \subfloat[Card piles for \faLayerGroup.]{
        \includegraphics[width=0.47\linewidth]{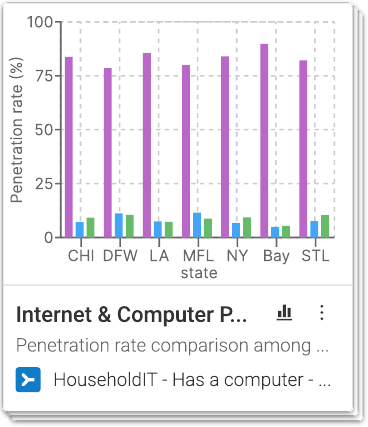}    
        \label{fig:chart-group}
    }
    \caption{\textbf{Visualization cards.}
    DrillVis uses \textit{cards} as containers for visualizations. 
    Single cards (left) represent a single chart atom~\faAtom{}; piled card visualization (right) represent piles~\faLayerGroup.
    }
    \label{fig:chart-panels}
\end{figure}

\subsection{Visual Interface}

The DrillVis interface consists of a \textit{drillboard view} (center of Figure~\ref{subfig:reader-mode-overview}) and a \textit{tree view} (left margin in Figure~\ref{subfig:reader-mode-overview}).
The drillboard view shows the current state of the drillboard. 
Each chart in the view is represented by a \textit{visualization card} (Figure~\ref{fig:chart-panels}).
Chart atoms~\faAtom{} are represented by a single card (Figure~\ref{fig:chart-atom}), whereas piles~\faObjectGroup{} are shown as card piles (Figure~\ref{fig:chart-group}).
The cards are laid out on the drillboard view using a regular grid that is organized from left to right, top to bottom.
The layout is continually rearranged to fill left and up; while this does lead to visual instability as cards are drilled down and rolled up, the order of the charts specified by the author is always maintained.

Three options for managing the drillboard space are possible.
One option is for cards to be a \textbf{fixed size}, which means that the grid is gradually filled out as the aggregation hierarchy is drilled down, and can eventually be scrolled to see all the charts.
Another option is to use a \textbf{space-filling layout}, where the visualization cards are dynamically resized to always fill the screen. 
This may lead to a situation where the cards become too small to view properly; in this situation, automatically rolling up distant cards into groups until the size is sufficiently large (similar to SpaceTree~\cite{DBLP:conf/infovis/PlaisantGB02}) is an option.
Finally, our current layout algorithm requires that all visualization cards have the same height, but does allow for resizing individual card widths.

The drillboard view supports multiple interactions. 
Hovering over a card representing a pile~\faObjectGroup{} highlights the pile, indicating that drill-down is possible.
Similarly, the cards for the parent group~\faObjectGroup{} are also highlighted, indicating what will happen if the user a rolls-up.
Furthermore, the card backgrounds are colored so that sibling cards have a consistent appearance to indicate their relation.
Finally, the visualizations inside each card can also be interacted with.

The tree view shows the current state of the aggregation hierarchy using a tree widget, similar to a file system browser.
The view will update as users drill down and roll up in the hierarchy.
Hovering over nodes in the view highlights the corresponding charts on the drillboard view.
Furthermore, the view itself can be interacted with to roll up and drill down into the hierarchy.


\subsection{Aggregation Operators}

We implement a part of the aggregation operations from Section~\ref{subsec:aggregation_operations}.
Note that the current implementation does not support the Label~\faFlag{} or Juxtaposition~\faPlusCircle{} operators.
Their addition would be straightforward.

\begin{figure}[!ht]
    \centering
    \includegraphics[width=\linewidth]{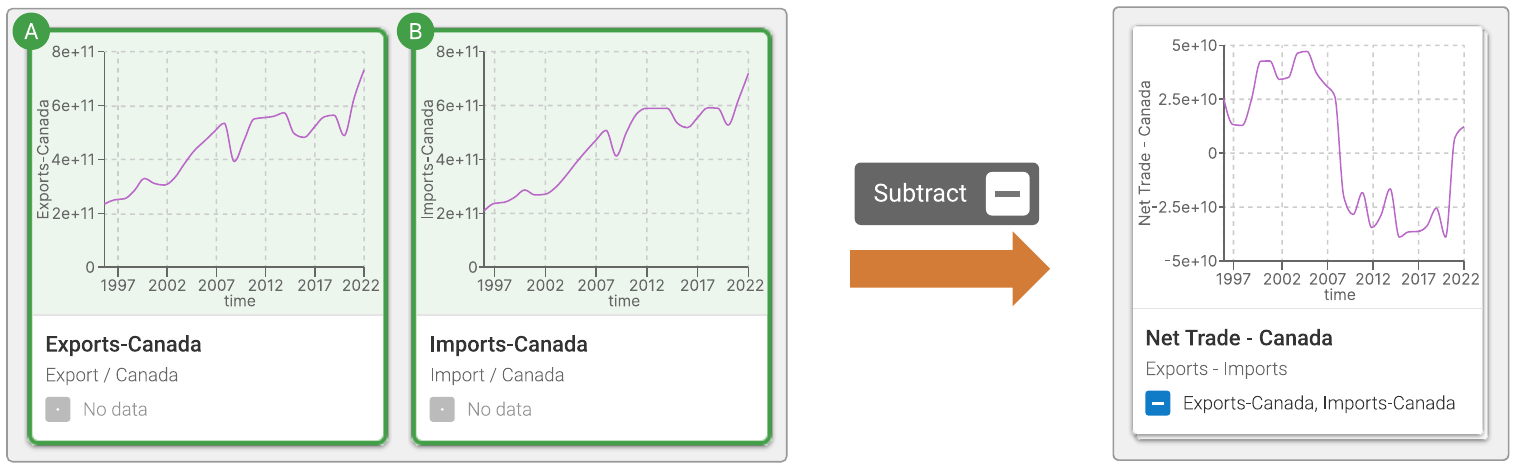}
    \caption{\texttt{Summarize~\faCalculator{} (subtraction).}
    Two charts for exports and imports are summarized using their difference.
    }
    \label{fig:summarize}
\end{figure}

\medskip\noindent\textbf{\faCalculator~\texttt{Summarize}} enables merging charts by performing direct calculations on the underlying datasets (Figure~\ref{fig:summarize}).
Supported calculations include addition, subtraction, multiplication, division, and averaging, each requiring compatibility in $y$-axis values across the selected charts.

\begin{figure}[!ht]
    \centering
    \includegraphics[width=\linewidth]{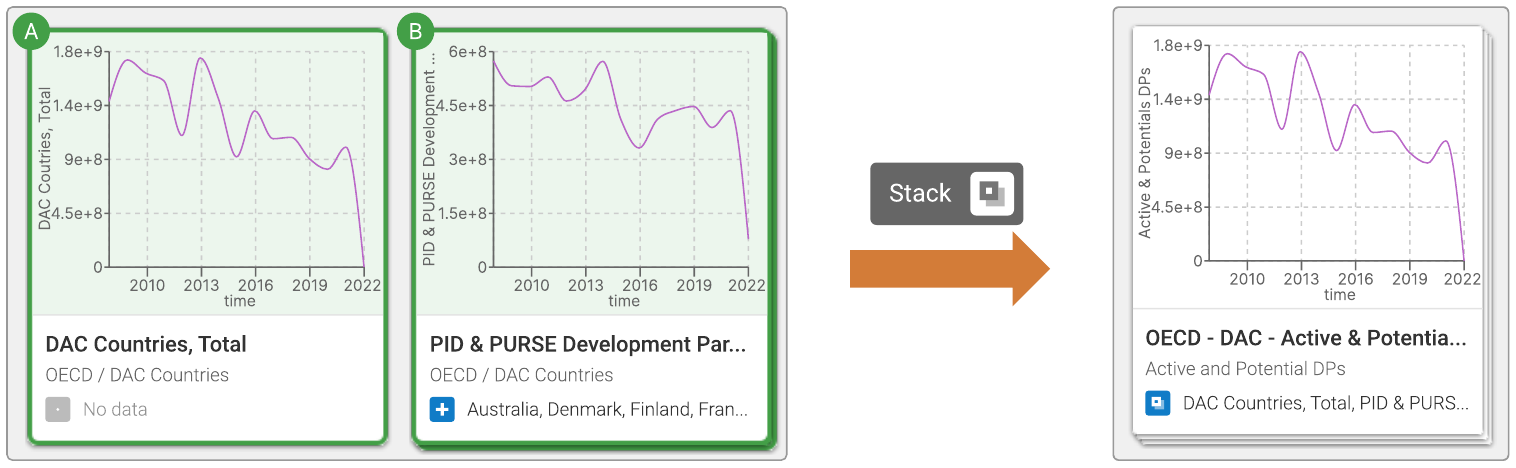}
    \caption{\texttt{Select~\faStar{}.}
    One of two merged charts are selected as archetype.
    }
    \label{fig:select}
\end{figure}

\medskip\noindent\textbf{\faStar~\texttt{Select}} merges multiple charts and lets the author choose a single one to represent them as an archetype (Figure~\ref{fig:select}).

\medskip\noindent\textbf{\faAngleDoubleDown~\texttt{Project}} generates a scatterplot from two data dimensions represented by child charts.
Note that our implementation currently does \textbf{not} support merging more than two chart into parallel coordinate plots.

\begin{figure}[!ht]
    \centering
    \includegraphics[width=\linewidth]{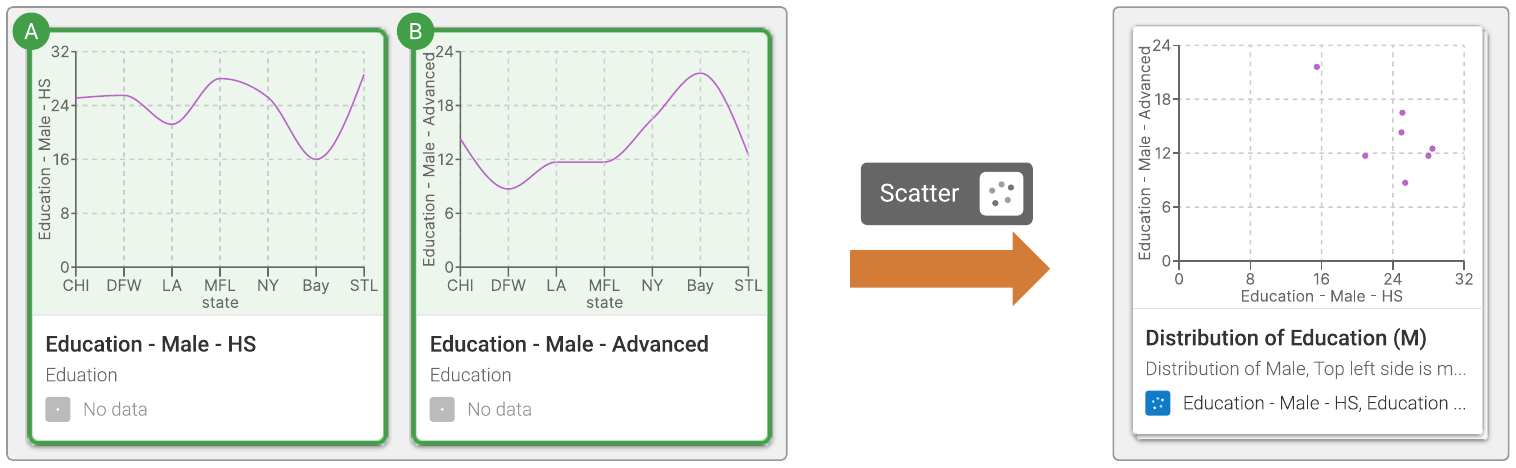}
    \caption{\texttt{Project~\faAngleDoubleDown.}
    Two data dimensions from charts are projected onto the axes of a scatterplot.
    }
    \label{fig:project}
\end{figure}

\medskip\noindent\textbf{\faLayerGroup~\texttt{Overlay}} superimposes charts in the same visual space~\cite{DBLP:conf/apvis/JavedE12}. 
The user can choose whether to merge or use two separate Y-axes.

\begin{figure}[!ht]
    \centering
    \includegraphics[width=\linewidth]{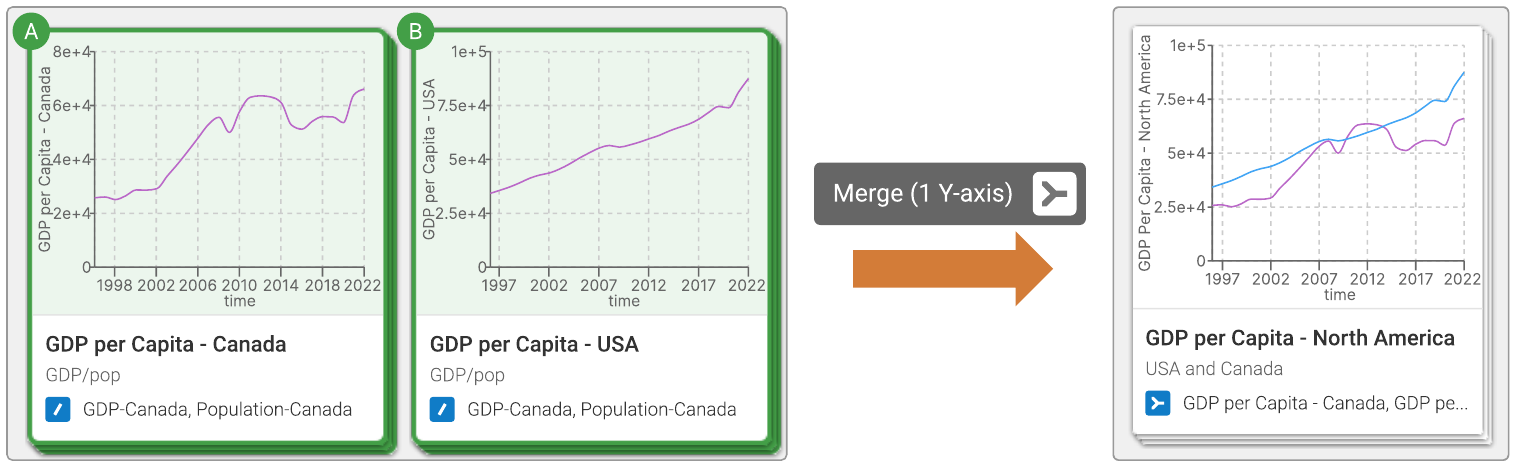}
    \caption{\texttt{Overlay~\faLayerGroup{}.}
    Two charts overlaid in the same visual space.
    }
    \label{fig:overlay}
\end{figure}

In addition, when selecting operations in the operation panel, users can also add a title and additional annotations to the chart (see Figure~\ref{fig:teaser}\textcircled{3}). 
They can split the charts, rename the title, and edit explanations from the treeview interface (see Figure~\ref{fig:teaser}\textcircled{2}).

\subsection{Author Mode}
\label{subsec:authoring-mode}

The authoring mode in DrillVis is designed to facilitate the construction of hierarchical visualizations (DG4).
This mode encompasses two primary functionalities: (1) \textbf{generating charts} from user-provided tabular datasets and (2) structuring the visualization hierarchy by \textbf{aggregating charts} using existing operations.

\medskip
\noindent\textbf{Generating charts.} DrillVis supports the ingestion of tabular data in formats such as CSV or TSV, accommodating both continuous and discrete data types.
For continuous data, such as time-series, the system automatically assigns time data to the $x$-axis and corresponding values to the $y$-axis, defaulting to a line chart presentation with an option to switch to a histogram.
Discrete data, representing categorical information, is visualized with categories on the x-axis and their values on the y-axis in the form of bar charts.
Authors can choose the type of chart between line and bar charts, or could opt for other charts. 
For example, authors can use scatterplots to understand the relationship between two features, as shown in Figure~\ref{fig:project}.

We assume that the dataset contains features in the first row and time tags in the first column. 
However, features themselves could belong to a group (e.g., cats and dogs are part of mammals), and a single row may be insufficient to hold all the dataset metadata. 
Even in such scenarios, the system automatically processes these relationships \addt{(DG4)}.

\begin{figure}[tb]
    \centering
    \frame{\includegraphics[width=\columnwidth]{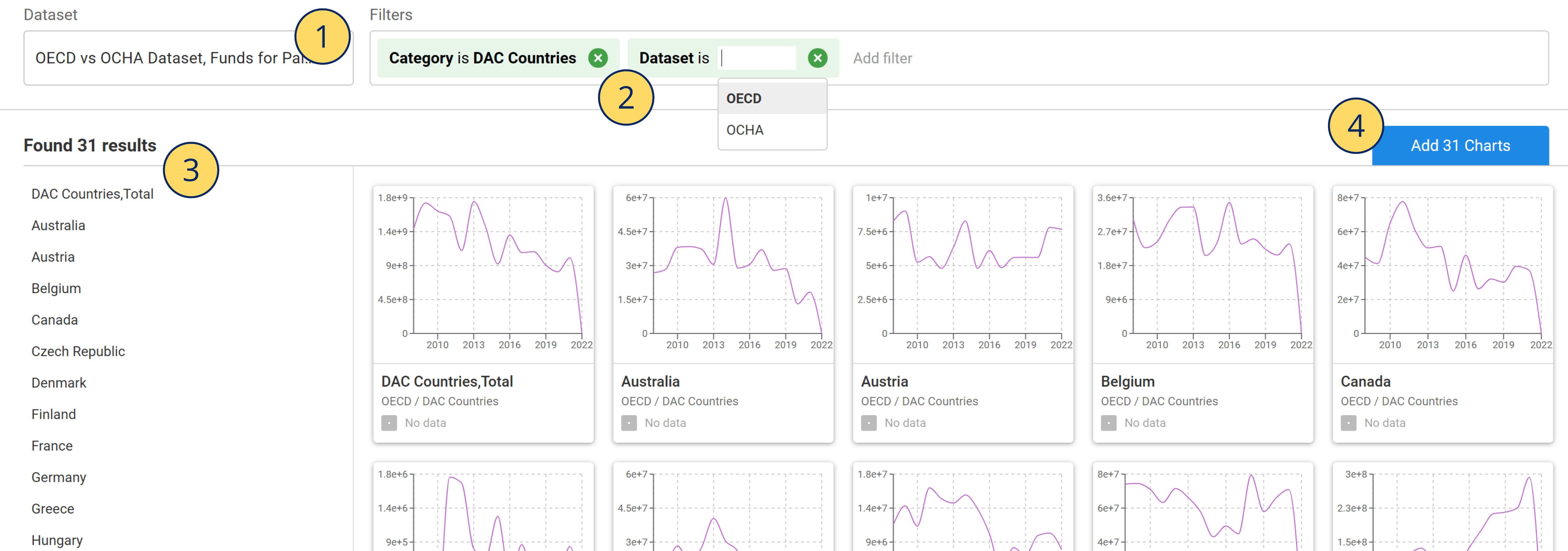}}
    \caption{\textbf{Chart selection in DrillVis.} Upon choosing a dataset ~\textcircled{1}, using multilevel dropdown menu authors can display charts of their interest, and choose those of their interest by clicking `add X charts' in~\textcircled{4}. The displayed charts are also shown in the treeview on \textcircled{3}.}
    \label{fig:author-data-query}
\end{figure}

In order to bring charts to the main view, authors can get access to the uploaded datasets via Figure~\ref{fig:author-data-query}~\textcircled{1}. 
Then, they can select groups and sub-groups of the dataset via a multilevel dropdown menu.~(see Figure~\ref{fig:author-data-query}~\textcircled{2}).
They can specify multiple sub-groups to specify charts of their interests.
For example, in a car dataset, it is possible to search for two conditions: cars that are SUVs and have a fuel efficiency of over 15 km/h.  
Then, clicking `Add X Charts' (Figure~\ref{fig:author-data-query}~\textcircled{4}) will display the selected charts in the main view.
Additionally, users have the capability to customize chart titles and annotate them with explanatory text.


\medskip
\noindent\textbf{Aggregating charts. }The aggregation process begins with the selection of charts to be merged (Figure~\ref{fig:teaser}(A)~\textcircled{1}), aiming to create a hierarchical structure of visualizations.
Consistent with Section~\ref{subsec:aggregation_operations}, DrillVis provides various merge operator implementations (Figure~\ref{fig:teaser}(A)~\textcircled{4}) for chart aggregation; see above.
When aggregating charts, there are rules to follow. 
First, charts from different datasets can only be combined if they have the same x-axis ranges or, for bar charts, the same x-axis values. 
\addt{We automatically disable inapplicable aggregation operations by graying them out.
(see Figure~\ref{fig:author-readermodes}(a)~\textcircled{3}) (DG4). 
Furthermore, the treeview interface is displayed at the left to facilitate checking of how the hierarchy is made (DG4).} 
Second, the system currently only supports tabular datasets and cannot yet visualize complex network graphs or advanced bespoke visualizations. 
We leave this as future work.

\medskip
\noindent\textbf{Defining views.} 
The design of DrillVis's authoring mode is underpinned by the assumption that users with varying levels of domain knowledge require different levels of detail and abstraction in visualizations (Figure~\ref{subfig:reader-mode-overview}~\textcircled{1}).
Thus, it allows authors to create and save multiple \textit{views} of the visualization hierarchy, catering to readers with different expertise levels.
This flexibility ensures that DrillVis can serve as a versatile tool for personalized data exploration and insight generation.
The bottom (all charts) and top (the fewest number of charts in the hierarchy) are automatically defined views representing the maximum and minimum levels of detail, respectively.
Once the authors have fully created the hierarchy of charts at their needs, they can save the model, and the hierarchy can be viewed in the reader mode.

\medskip
\noindent\textbf{Constraints in aggregation. }
\addt{For many operations, certain constraints must be met. 
For example, calculations such as averages, sums, or differences are only possible when both the x- and y-axes of the charts share the same dimensions. 
For projection, the x-axis of both charts must share the same dimension. 
Multiple lines can be overlaid in a single chart if the two charts also share the same dimensions on the y-axis.
}

\addt{Even though we used relatively simple charts, we encounter numerous scenarios where the merging process is not applicable.
For example, consider a chart that is already merged as a juxtaposition, alongside another chart that overlays two visualizations. 
In this situation, we identify a few operations that are not possible: arithmetic summarization, projection, and additional overlays. 
Juxtaposition may still work, but only until the unit charts become too small to be legible.
Another example involves aggregating two charts, one of which is merged through text labeling, and the other through projection. While aggregation through labeling or an archetype might be feasible, many arithmetic operations or overlays are not.
In our case, we have automatically disabled such merge operations.}

\medskip
\noindent\textbf{Consistency in aggregation policies. }
\addt{We may face confusion when the meaning of these aggregations is ambiguous. 
For instance, should we allow the arithmetic merging of a leaf line chart with another line chart previously merged as an `archetype' using two line charts and one labeled chart? 
In our case, we decided against this operation because the resulting aggregation would be difficult to interpret.}

\addt{We do not claim that our approach is the only solution. 
Depending on the goals of Drillboards, policies may vary. 
We argue that consistency is crucial to avoid confusing users, so well-crafted guidelines are needed to determine when merging is permitted and when it is not.}





\subsection{Reader Mode}
\label{subsec:reader-mode}

DrillVis's reader mode is designed to facilitate the exploration of visualization hierarchies created by authors to match varying levels of user expertise.
This mode assumes that the reader first chooses a pre-defined view corresponding to their expertise level (DG2, DR3).
Navigation within this structured hierarchy is enabled through two core mechanisms: (1) \textit{drill-down} and (2) \textit{roll-up}~\cite{Elmqvist2010}, allowing users to navigate between different levels of detail within the dashboard.

\medskip\noindent\textbf{Drill-down. }piles~\faObjectGroup{} containing child elements are identified using a visual cue resembling a stacked card structure beneath each chart (Figure~\ref{fig:teaser}(B)~\textcircled{2}).
To increase the level of detail by expanding such a group, the reader can initiate a \textit{drill-down} by clicking on the group.
This action highlights the chart, which then animates to reveal its constituent child charts, providing a visual transition that aids in tracking the exploration path.
Each subsequent level of drill-down is visually distinguished through increased opacity and color differentiation for up to five selectable sub-trees, with each new child chart labeled with a number indicating its depth within the hierarchy.
The presence of additional drill-down options is signified by remaining stacks beneath the chart or can be inferred from icons and titles at the card bottom (Figure~\ref{fig:teaser}(B)~\textcircled{6}).

Concurrent with each drill-down action, the treeview interface on the dashboard's left side updates to reflect the hierarchical structure being explored (Figure~\ref{fig:teaser}(B)~\textcircled{1}$\rightarrow$\textcircled{3}).
This treeview provides an overview of the chart hierarchy, highlighting active subtrees in synchronization with the main dashboard's color scheme and distinguishing between visible and hidden charts through bold and grayed-out titles, respectively.

\medskip\noindent\textbf{Roll-up. }The roll-up function acts as the counterpart to drill-down, enabling users to navigate back to higher levels of the visualization hierarchy (Figure~\ref{fig:teaser}(B)~\textcircled{5}).
This process is initiated by clicking on a chart at the deepest explored depth with the right mouse button, triggering a roll-up to the preceding level of abstraction.
As the visualization structure condenses, the treeview interface mirrors this action by collapsing child nodes, thereby synchronizing the hierarchical view with the main dashboard's state (Figure~\ref{fig:teaser}(B)~\textcircled{6}$\rightarrow$\textcircled{2}). 
This bidirectional navigational approach ensures that readers can fluidly transition between detailed and abstract views, facilitating a comprehensive understanding of the visualized data's underlying structure and relationships.

\subsection{Implementation Notes}
\label{subsec:implementation}

We implemented Drillboards in TypeScript using React\footnote{\url{https://react.dev/}} and Node.js.
The user interface is built and designed manually by the authors using the Emotion library.\footnote{\url{https://emotion.sh/}} 
We also further customized components (i.e. input, menu) using MUI.\footnote{\url{https://mui.com/}}
The charts are generated using Recharts.\footnote{\url{https://recharts.org/}}
We used Cloudflare R2 for cloud storage.

\subsection{Usecase Example}
\label{sec:usecase_Senario}

We provide a drillboard example using water sensor data to explain how students on a field trip could use the tool to filter and check for farm anomalies.



\begin{figure}[htb]
    \centering
    \includegraphics[width=\columnwidth]{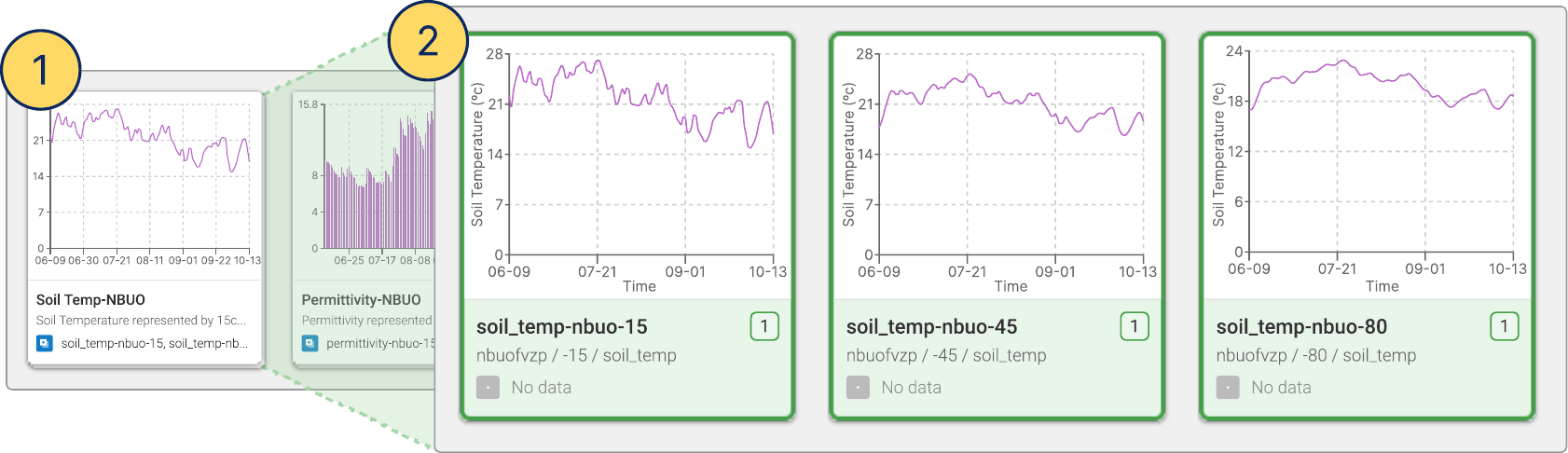}
    \caption{\textbf{DrillVis showing water sensor data.} The following operation is an example of \textit{select}, where one of the merged charts are shown as an archetype. \textcircled{1} represents the original chart, and \textcircled{2} depicts its expanded form after left-clicking on \textcircled{1}.}
    \label{fig:farmdata}
\end{figure}

Bob, an expert in agriculture, wants to monitor water sensors on a farm located in Maryland, USA. 
He receives a time-series dataset about the farm he is located, and needs to instruct incoming interns about how to monitor water sensor data effectively. 
To ensure efficiency, he uses DrillVis to educate interns on key features to analyze in the water sensor data.
To that end, he first builds a version for novice users to educate and instruct interns.
In doing so, he uses the `select' operator to indicate the important feature, and group those that exhibit similar patterns. 
He also creates an `expert' mode, a version that contains more charts, with less abstraction.
An illustration of `select' is shown in Figure~\ref{fig:farmdata}.

\addt{When new interns arrive, Bob instructs them on monitoring soil conditions. He begins with key features at the top of the hierarchy using novice mode, then drills down (as shown in Figure~\ref{fig:farmdata}) to detail soil data. Novice users start with basic information, but as they gain experience, they must be ready to review more detailed data.}

\addt{After guiding novice interns, Bob instructs seasoned interns using the same drillboards hierarchy. For them, he selects expert mode to explain advanced features.}

\section{User Study}
\label{sec:results}

To evaluate the effectiveness of the drillboards technique, we recruited three domain experts who utilize data in their daily occupations to use it author a drillboard using the DrillVis prototype.
We then evaluate the effectiveness of the resulting drillboards with a group of casual end-users.
Here we describe the method and results from our study.

\begin{table}[htb]
    \centering
    \caption{\textbf{User study demographics.}
    Three domain experts from different backgrounds participated in the user study.}
    \scalebox{0.935}{
    \begin{tabular}{llllp{2.75cm}l}
        \toprule
        \textbf{{ID}} &
        \textbf{{Gender}} &
        \textbf{{Age}} &
        \textbf{{Education}} &
        \textbf{{Job Title}}&
        \textbf{{Experience}}\\
        \midrule
        \rowcolor{LightSteelBlue!30}
        P1 & Male & 33 & M.A. & Officer at international financial institution & 5 years\\
        P2 & Male & 33 & B.S. & Deputy director in governmental ministry & 10 years\\
        \rowcolor{LightSteelBlue!30}
        P3 & Male & 49 & Ph.D. & Assistant professor of statistics & 15 years \\
        \bottomrule
    \end{tabular}
    }
    \label{tab:exp_participant}
\end{table}

\def\sel{\texttt{sel}\xspace}
\def\sum{\texttt{sum}\xspace}
\def\add{\texttt{add}\xspace}
\def\sub{\texttt{sub}\xspace}
\def\div{\texttt{div}\xspace}
\def\ovl{\texttt{ovl}\xspace} 

\begin{table*}[t]
    \centering
    \caption{\textbf{How experts used drillboards.}
    Here is the statistics of how domain experts created their hierarchies using drillboards in two different modes, novice and expert. 
    \sel, \sum, \ovl, \add, \sub, and \div represent select, summarize, overlay, add, subtract, and divide, respectively.}
    \begin{tabular}{clllllll}
    \toprule
         &  \multirow{2}{*}{\raisebox{-2.5pt}{\makecell[tl]{\textbf{\# of base} \\ \textbf{charts}}}} & \multicolumn{2}{l}{\textbf{Final \# of charts}} & \multicolumn{2}{l}{\textbf{Final \# of ops.}} & \multicolumn{2}{l}{\textbf{Type and frequency of the used operators}}\\
         \cmidrule(lr){3-4} \cmidrule(lr){5-6} \cmidrule(lr){7-8}
         &    & novice & expert & novice & expert & novice & expert \\
         \midrule
     \rowcolor{LightSteelBlue!30}P1  &  96 &  8 & 26 & 26 & 22  &  15 \sel , 11 \sum ($=$ 11 \add) & 11 \sel , 11 \sum ($=$ 11 \add)  \\
     P2  & 32 & 6 & 11 & 21 & 14 &  19 \sum (=8 \add $+$ 6 \sub $+$ 5 \div), 3 \ovl  & 11 \sum (=2 \add $+$ 6 \sub $+$ 3 \div), 3 \ovl  \\
     \rowcolor{LightSteelBlue!30} P3 & 9 & 3 & 9 & 3 & 0  & 3 \sel & None \\
     \bottomrule
    \end{tabular}
    \label{tab:results-drillboards-use}
\end{table*}

\subsection{Participants: Experts and End-Users}

Our study had two separate groups of participants---\textbf{domain experts} and \textbf{casual end-users}---who were engaged at separate phases of the experiment.
The domain experts were three individuals who use data in their daily lives; Table~\ref{tab:exp_participant} gives an overview.
For the casual end-users, we recruited 10 participants with no special relevance in topics mentioned by our domain experts. 

\subsection{Procedure}
\label{subsec:experiment-results}

Our study consisted of four phases, of which the first three involved our expert participants and the last one involved our casual end-users. 
All study sessions were conducted via Zoom videoconferencing. 
Participants shared their screen and were audio and video recorded.
Our study was approved by our institution's ethics review board.

\medskip\noindent\textbf{Phase I: Initial Interview. }
We started the study by meeting with each of our domain experts to collect their informed consent, gather basic demographic information, and interview them about their data analytics expertise.
We also demonstrated the reader view in the DrillVis tool with the farm dataset and allowed them to use it while following a think-aloud protocol.
The purpose was to familiarize the experts with drillboards and instruct them for the authoring task in the next meeting. 
Finally, the experts were asked to produce a dataset consisting of at least 3-4 data tables to use for the next phase.
This session lasted approximately 20 minutes.

The experts provided us with the following datasets:
\begin{itemize}
    
    \item\textbf{Source of funds granted to a country.}
    The officer from the international financial institution wanted to compare the datasets about funds going into a country of two organizations: the Organisation for Economic Co-operation and Development~(OECD), and United Nations Office for the Coordination of Humanitarian Affairs~(OCHA).
    
    \item\textbf{The scarcity of employees in a field.}
    The deputy director of a governmental ministry wanted to investigate on workforce shortages in a designated field. 
    To support this investigation, he provided various datasets covering the sector across different regions, including details like years of experience and specific skills.
    
    \item\textbf{Regional household data for a country.}
    The assistant professor of statistics aimed to examine the disparities between wealthy and impoverished households across different regions of the country.
    He presented a dataset encompassing various features, such as home ownership, internet subscriptions, educational levels, the age of properties, and more for this analysis.
\end{itemize}

\medskip\noindent\textbf{Phase II: Data Preparation. } In this phase, the research team adapted the datasets provided by the experts so that they could be loaded into the DrillVis tool. 
While DrillVis nominally accepts any CSV file, there typically was some formatting and curation in this phase. 
For example, one participant provided an Excel spreadsheet with multiple worksheets; we split these into separate CSV files to use in the DrillVis tool.
This phase lasted 3-4 days on average. 
\addt{The intent of giving 3 to 4 days was because these domain experts were extremely busy. 
It is not because it took 3 to 4 days, but we wanted them to do the study whenever they had time. 
Furthermore, the goal was to give them time to think of the data before designing their own personalized structure.}

\medskip\noindent\textbf{Phase III: Drillboard Authoring. } When we met with our domain experts, we first demonstrated the authoring mode in the DrillVis tool using the farm dataset that they had seen in Phase I.
During this demonstration, we showed all the aggregation and layout operators. 
We then let the experts themselves try out the tool and ask questions. 
Once they indicated they were ready to proceed, we provided them with a link to the DrillVis website with their dataset preloaded.
We now asked the experts to use the authoring mode to create a new drillboard for their own data.
We instructed them to author the drillboard so that it had two levels of detail that would be suitable for an expert and for a complete novice.
We also asked them to list three in-depth questions about the dataset that should be answerable using the drillboard.
Again, the experts followed a think-aloud protocol.
There was no time limit during this phase, and we did not collect any metrics beyond observational notes and recordings.
Participants were allowed to ask questions freely about the DrillVis tool.
These sessions lasted between 1 hour to 1h and 30 minutes, and ended when the expert was satisfied.

\medskip\noindent\textbf{Phase IV: Drillboard Analysis. } We finally met with casual end-users who we had carefully screened to have no prior specialized knowledge about the three datasets provided by our expert users.
We collected their informed consent, presented them with the basic objective of the study, and demonstrated drillboards using the farm dataset.
Participants were then allowed to explore this drillboard in their own browser while the evaluator answered their questions. 
Over the duration of approximately forty minutes to one hour in total, these participants were then given access to each of the drillboards created by our domain experts.
During this time, end-users were allowed to freely navigate the aggregation hierarchy or to use the pre-defined views.
They all spent approximately 10 minutes per drillboard attempting to answer the questions prepared by the experts.
During this time, we asked the end-users to verbalize their reasoning process, and answer each question as a full sentence, at least. 
We closed the session by interviewing the participants using a semi-structured interview format designed to glean their perception and feelings about DrillVis.








\subsection{Usability Study}

\begin{figure}[tb]
    \centering
    \includegraphics[width=\columnwidth]{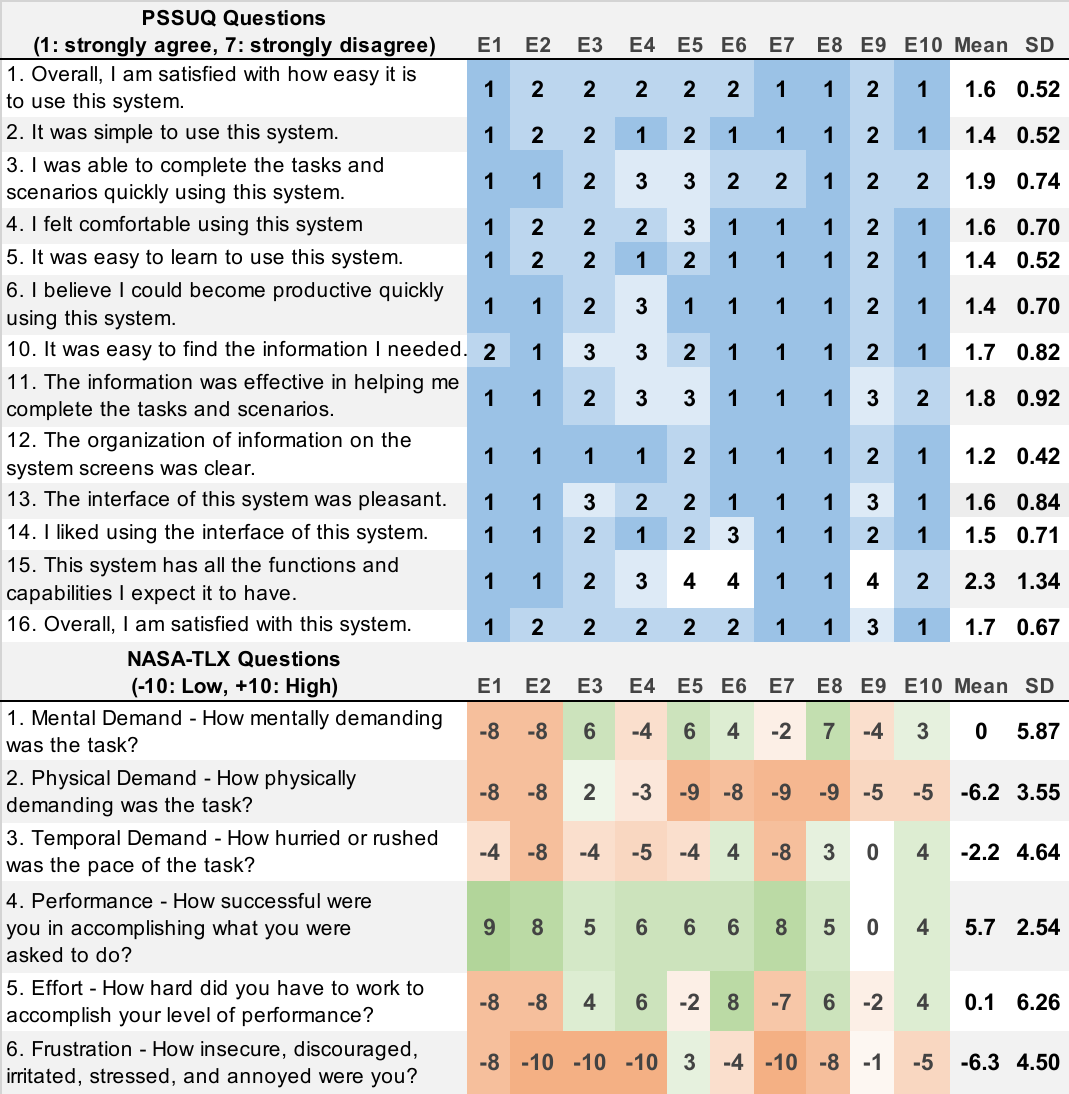}
    \caption{\textbf{Drillboard usability study.} \addt{We conducted usability tests using the PSSUQ and NASA-TLX questionnaires to evaluate user satisfaction and workload in Drillvis. The Mean indicates the arithmetic mean scores of all participants, and SD refers to their standard deviation.
    Three PSSUQ questions (7, 8, and 9) were excluded as they were not relevant to our tool. 
    Results indicate that participants were generally satisfied, and NASA-TLX findings suggest the workload was manageable, though some found the task slightly mentally demanding.} }
    \label{fig:usability}
\end{figure}

\addt{As a post-study, we conducted a usability evaluation with participants using the PSSUQ and NASA-TLX questionnaires to assess satisfaction and workload in Drillvis.
Figure~\ref{fig:usability} presents the results of this usability study.
We excluded three PSSUQ questions that were not relevant to our system’s conditions: e.g., `\textit{7. The system gave error messages that clearly told me how to fix problems},' `\textit{8. Whenever I made a mistake using the system, I could recover easily and quickly}'), and `\textit{9. The information (such as online help, on-screen messages, and other documentation) provided with this system was clear.}'}

\addt{Overall, participants were satisfied with using Drillvis, as reflected in the PSSUQ results.
Most questions had an average score between 1 and 2, with little variation in standard deviation.
The highest standard deviation was observed in Q15, which asked whether the system had all the expected functions and capabilities. 
Responses to this question varied slightly, though none of the participant expressed disagreement.}

\addt{In the NASA-TLX questionnaire, participants generally did not find the workload overwhelming.
However, responses regarding mental demand and effort were mixed, with some participants reporting a substantial cognitive load. 
This suggests that while the system was manageable for most, certain aspects required significant mental effort for some users.
We believe this may be linked to the issue of visual stability.
We further discuss this in Sec.~\ref{subsec:limitations}.}

\subsection{Results}

Here we report on our overall results from this qualitative multi-phase experiment. 
We first present the drillboards produced by the domain experts. 
We then report on our observations from the domain experts, think-aloud results, and from post-study interview. 
This is followed by the same data for our casual end-users.

\subsubsection{Authored Drillboards}

\begin{figure}[tb]
    \centering
    \includegraphics[width=\columnwidth]{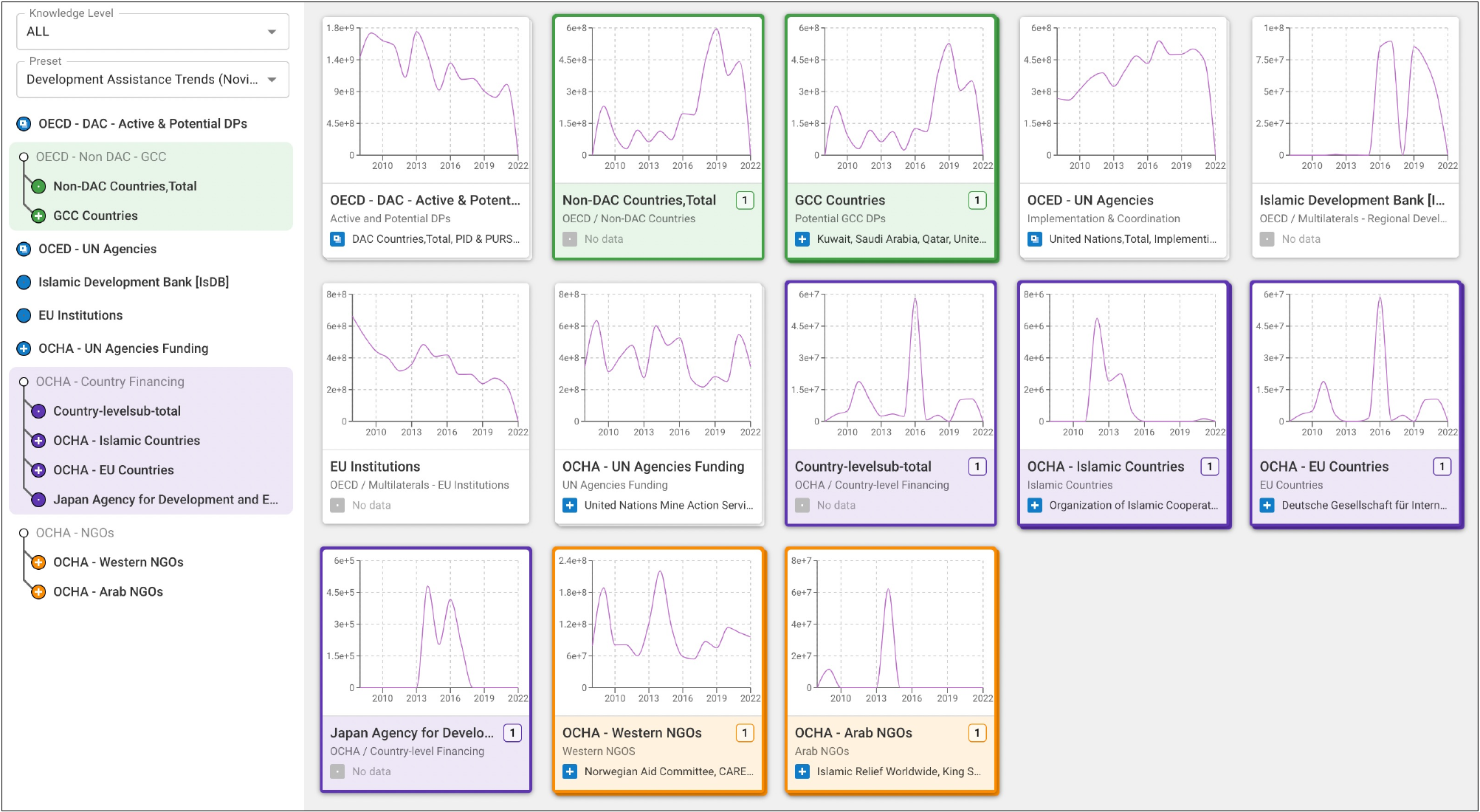}
    \caption{\textbf{P1's drillboard.}
    P1 is analyzing financial assistance to a nation using two data sources.
    P1's representation, shown above, is tailored for novices with greater abstraction for easier understanding.}
    \label{fig:P1-expert}
\end{figure}

We describe the drillboards produced by the domain experts. 
An example of drillboards authored by the expert P1 is shown in Figure~\ref{fig:P1-expert}.
Also, Table~\ref{tab:results-drillboards-use} displays the quantity of charts, and the quantity/type of operators the experts used to create the hierarchies of novice and expert modes using drillboards.


P1 deployed 96 leaf chart atoms to construct his hierarchy. 
He used the `add' operator multiple times to merge countries or NGOs with small contributions and frequently used the select' operator, choosing the chart with the largest value to represent grouped charts. 
In the end, he created a hierarchy with 8 chart atoms for novice users and presented 26 chart atoms for expert users.
P2 deployed 32 leaf chart atoms, focusing on arithmetic operations (5 \div, 8 \add, 6 \sub) and using `overlay' several times. 
This resulted in 6 final charts for novices and 11 for experts.
P3 used 9 leaf chart atoms, summarizing by selecting the chart with the most prominent trend from each group. 
He used the `select' aggregation three times to formulate his hierarchy, resulting in 3 charts for novice mode and all 9 charts for expert mode.

\subsubsection{Domain Experts}

Here we review the findings of how experts authored using drillboards based on our observations and post-study interview, insights gained on the utilization of drillboards by experts, informed by our observations and interviews conducted after the study.

\medskip\noindent\textbf{Expert P1.} \textit{Intent:} P1 aimed to examine numerical and trend discrepancies between two datasets on financial sources provided to a specific country.
\textit{Process:} The datasets are, as mentioned in Section~\ref{subsec:experiment-results} from OECD, and OCHA. 
The OECD dataset contains financial support given by countries and international organizations from 2008 to 2022. 
Across all charts, the $x$-axis represents the year and the $y$-axis indicates the funds allocated. 
He initially loaded all potential visualizations from the dataset and discarded those not aligned with his goal. 
Using the addition operator, he aggregated small-scale funds into groups according to his standards, appending `OECD' to signify their origin. 
Similarly, he grouped small-scale funds from NGOs and UN-related organizations. 
Occasionally, he was surprised to find visualizations conflicting with his prior knowledge.

Unlike the OECD dataset, which covers country-based funding, the OCHA dataset targets financial inputs from global entities, including the UN and NGOs like Oxfam. 
P1 arranged the OCHA visualizations into groups reflecting his interests. 
Once the setup was finished, he scrutinized dataset differences.

\medskip\noindent\textbf{Expert P2.} \textit{Intent:} P2 aimed to understand the magnitude of an industry's labor shortfall nationwide. 
As a government official, his priority was to quickly grasp the situation and identify problem areas.
\textit{Process:} He analyzed the progression and anomalies in features like the count of skilled personnel and population per region from 1995 to 2022. 
He focused on regions with trends differing from the overall country's trends. 
In his charts, the x-axis showed time, while the y-axis reflected various metrics like expert/novice counts or employee numbers by company size. 
He extensively used arithmetic operations, including addition, subtraction, and division. 
He experimented with these and overlay operations to understand the data before finalizing drillboards for both expert and novice modes.

\medskip\noindent\textbf{Expert P3.} \textit{Intent: }P3 aimed to study wealth disparity among cities in different areas of a country. 
His dataset included 45 wealth indicators, such as internet access and education levels, for 7 cities across diverse regions.
\textit{Process: }In all of his charts, the x-axis displayed the 7 cities, while the y-axis showed values for each feature in the dataset. 
P3 initially uploaded all charts to the main dashboard and deleted unnecessary ones. 
At first, he created hierarchies by applying addition and subtraction operators to merge data. 
After a brief hesitation, he split them again, and used `select' operator to represent similar groups with one chart, believing a single representative chart would convey the message. 
He saved versions for both novices and experts.


\subsubsection{Casual End-Users}
We summarize findings from 10 casual end-users based on observations and post-study interviews. 
Each drillboard had 3 questions provided by the authors, with a 10-minute soft deadline to ensure consistency. 
Authors created 3 short-answer questions per set to convey their messages, solvable within two minutes each using DrillVis. 
To ensure ample time without rushing, we limited tasks to 10 minutes per drillboard.
For end-users, we instructed them to provide not just short replies but also a one-sentence justification for their responses.
Table~\ref{tab:author-time-results} shows the demographics of 10 casual end-users who participated in the experiment. 
It also shows time taken to solve questions from 3 drillboards using DrillVis.

All end-users were able to complete the questions of the three different datasets promptly---9 out of 10 not exceeding 10 minutes---except one participant, E9, who ended in 10 minutes and 24 seconds.
Note that the average time taken to solve 3 questions for P1, P2, and P3 were 8m 39s, 7m 56s, and 7m 49s. 
All casual end-users were able to get the solutions correct and provide reasons behind them.
We observed that end-users exhibited a preference for the novice mode over the expert mode, consistently choosing it even for P3's visualization, which contained only 9 charts in the most abstracted version of expert mode and 3 in the novice mode.
While P1 and P2 showed mixed use of novice and expert modes, all participants used novice mode in P3's example.
The questions and answers provided by the end-users can be found in~\url{https://osf.io/qfmbj/}.

\begin{table*}[!htb]
    \centering
    \caption{\textbf{Demographics and completion time for end-users (E1-E10).}
    10 participants conducted the study as casual end users who have never seen the system, nor the dataset before. 
    We present the time it took casual end-users three questions from three experts, respectively.
    To each participants' customized drillboards, we provided 10 minutes to solve three questions.
    Furthermore, when asked whether to use novice or expert mode, all end-users preferred to use novice mode. }
    \begin{tabular}{llllllllc}
        \toprule
        \textbf{{ID}} &
        \textbf{{Gender}} &
        \textbf{{Education}} &
        \textbf{{Occupation}} &
        \textbf{{Age}} &
        \textbf{{P1's drillboard}} &
        \textbf{{P2's drillboard}} &
        \textbf{{P3's drillboard}} &
        \textbf{{Novice vs. expert mode}} \\
        \midrule
        E1 & Male & M.S. & Research scientist & 32 & 8m 30s & 6m 52s & 5m 14s & Novice \\ 
        \rowcolor{LightSteelBlue!30}
        E2 & Male & M.S. & IT engineer & 30 & 8m 32s & 9m 13s & 8m 31s & Novice\\  
        E3 & Male & Ph.D. & Research scientist & 32 & 8m 34s & 4m 17s & 4m 32s & Novice\\ 
        \rowcolor{LightSteelBlue!30}
        E4 & Male & M.S. & Research scientist & 27 & 9m 15s & 9m 29s & 9m 33s & Novice\\  
        E5 & Male & B.S. & IT engineer & 33 & 9m 25s & 9m 18s & 7m 15s & Novice\\  
        \rowcolor{LightSteelBlue!30}
        E6 & Male & M.S. & Semiconductor engineer & 31 & 3m 21s & 8m 55s & 9m 07s  & Novice\\ 
        E7 & Male & M.A. & Financial consultant & 32 & 9m 37s & 4m 35s & 9m 48s & Novice\\ 
        \rowcolor{LightSteelBlue!30}
        E8 & Female & B.A. & Marketing analyst & 29 & 9m 46s & 9m 53s & 6m 21s & Novice\\ 
        E9 & Male & Ph.D. & Research scientist & 33 & 9m 53s & 10m 24s & 9m 45s & Novice \\ 
        \rowcolor{LightSteelBlue!30}
        E10 & Male & B.S. & Research scientist & 26 & 9m 46s & 6m 21s & 8m 04s & Novice\\ 
        \bottomrule
    \end{tabular}
    \label{tab:author-time-results}
\end{table*}

\section{Discussion and Limitations}
\label{sec:discussion}

Here we discuss the outcomes of our study, generalizing the insights gleaned from the deployment of DrillVis and examining its implications for the broader field of adaptive visualization dashboards.
Our analysis reveals key findings regarding user interaction with hierarchical data representations, personalization of data exploration, and the overall utility of drillboards in accommodating diverse user needs.
We also identify limitations inherent to our work and propose directions for future research. 

\subsection{Explaining the Results}
\label{subsec:results-explain}

We explain the results from two parts: observation from experts (authors) and casual end-users (readers). 

\smallskip\noindent\textbf{Observations from experts.}
Each expert aggregated their hierarchy with different intent. 
P1 used the tool for macro-level trend analysis by aggregating many charts for broader insights. 
P2 focused on using arithmetic operations to quickly detect issues. 
P3 highlighted differences between wealthy and impoverished regions by emphasizing a key feature.
This suggests that progressive abstraction actually mirrors many expert's mental models of such data; humans have a tendency to summarize and abstract, and drillboards support this model.

The feedback from experts was generally positive. 
They praised the tool's ability to generate charts with increasing levels of abstraction, which effectively delivers their message (P1, P2, P3), and its facilitation of predefined views. 
P2 and P3 found chartwise operations more intuitive than tables and sentences, as graphical explanations better communicate data.

\smallskip\noindent\textbf{Observations from casual end-users.} First, participants took less than 10 minutes and answered correctly, showing they adapted swiftly to the interface. 
Second, the preference for novice mode with more chart units suggests that author guidance helps end-users understand authors' intents in large data sets.
Third, the time taken to solve the three questions is consistent despite the varying number of chart units. 
This indicates that hierarchical aggregate structure helps people efficiently understand the context and the message in the data.

According to the feedback from end-users, they were more optimistic about novice mode because the data was unfamiliar (E7, E8, E9) and large datasets required time to understand. 
Guidance helped quickly grasp content and issues (E1-5, E10).

\subsection{Limitations}
\label{subsec:limitations}


\textbf{Visual stability.} End-user challenges with perceiving and navigating the drillboard may be explained by the lack of visual stability.
The dynamic expansion and collapse of charts during drill-down and roll-up operations sometimes led to confusion, likely because the changing order and appearance of charts disrupted the participants' mental map of the data.
After trying three different versions of drillboards, two participants, E2, and E7, admitted confusion in when drilling down into various types of aggregate operators. 
E2 said, \textit{`The drill-down interaction is occasionally unpredictable; at times, it displays the data elements, sometimes decomposing a multi-line chart into individual lines, and at other times, presenting arithmetic operation elements. This lack of consistency leads to confusion.'}

In general, visual stability and the preservation of the user's mental map is a fundamental design aspect of visual representations; for example, it was one of the key shortcomings of the hyperbolic tree~\cite{DBLP:conf/chi/LampingRP95}.
Obviously, by freezing a drillboard for the end-user, the problem is eliminated.
However, drilling down and rolling up is central to the technique, so further design interventions are necessary to address this.
We may consider better highlighting, improved layout techniques (perhaps using space-enclosure), and careful animated transitions as possible solutions.

\textbf{Drillboards with different charts.} 
\addt{While our approach is adaptable to different chart types, each type has unique aggregation conditions and functionalities that require tailored policies. 
For instance, a treemap---which is inherently hierarchical---could be constructed using Drillboards, but the way it summarizes data (e.g., grouping levels) differs from how line or bar charts aggregate information. 
Certain interactions, such as overlay or projection, may also be less applicable. 
Similarly, in node-link diagrams, operations like labeling, summarization, and archetyping seem feasible, but projection and overlay may not work as intuitively. 
These differences indicate that Drillboards’ aggregation model is not universally applicable without modification. 
Therefore, while Drillboards can support various visualizations, each implementation requires specific adaptations, which we leave as future work.}

\subsection{Future Work}
\label{subsec:futurework}

We see potential for using drillboards for different purposes. 
First is as a chart-wise operational analytical tool. 
How effective are drillboards when chart-wise operations in author mode are used analytically? 
Comparative studies with existing data exploration tools are needed to evaluate their effectiveness. 
Another possibility is as a situated presentation tool for data-driven storytelling~\cite{Javed2013, DBLP:journals/cgf/MathisenHKGE19}. 
Delivering messages through chart-wise operations could prove an effective communication mechanism.
P2, who tried to deliver his analytical reasoning to the audience, said, \textit{``If I could use the authoring tool on-the-fly as a presentation, then I could easily explain to my audience the steps I took to get the end result.''}

We also identify methods to improve drillboards. 
One direction is to make them easier to use for a wider audience by providing effective guidance or automating the aggregation process. 
Furthermore, there are various charts that drillboards do not yet provide and numerous aggregation mechanisms that are not implemented.
All these issues are left for future work, however.

\section{Conclusion}
\label{sec:conclusion}

We have introduced drillboards, a novel adaptive visualization technique, and detailed its implementation within the DrillVis system.
Through illustrative case studies and a rigorous four-phase user study, we have validated the effectiveness of DrillVis in facilitating data exploration across different levels of user expertise: both experts and casual end-users.
This approach allows for a personalized interaction with data, offering various levels of detail and abstraction.
Our findings suggest that DrillVis significantly improves user engagement and understanding of complex datasets, potentially due to its capacity to tailor visual representations to individual user needs and cognitive styles.

\section*{Acknowledgments}
\noindent This work was supported partly by Villum Investigator grant VL-54492 by Villum Fonden.
Any opinions, findings, and conclusions expressed in this material are those of the authors and do not necessarily reflect the views of the funding agency. 
We would like to thank the anonymous reviewers for their feedback on this paper. 
We thank Sungwoo Jeon, Jisik Woo, and Eric Newburger for their comments on our work, as well as Hyeon Jeon for his comments on our illustrations.

\bibliographystyle{abbrv-doi}
\bibliography{drillboards}

\begin{thebibliography}{10}

\bibitem{alzoubi21educdash}
D.~Alzoubi, J.~Kelley, E.~Baran, S.~B.~Gilbert, A.~Karabulut~Ilgu, and S.~Jiang.
\newblock {TeachActive} feedback dashboard: Using automated classroom analytics to visualize pedagogical strategies at a glance.
\newblock In {\em Extended Abstracts of the {ACM} Conference on Human Factors in Computing Systems}. {ACM}, {New York, NY, USA}, 2021. doi: {{%
10\hspace{.1pt}\discretionary{.}{%
}{.}\hspace{.4pt}1145\discretionary{/}{%
}{/}3411763\hspace{.1pt}\discretionary{.}{%
}{.}\hspace{.4pt}3451709}}


\bibitem{bach23dashboardpatterns_tvcg}
B.~Bach, E.~Freeman, A.~Abdul-Rahman, C.~Turkay, S.~Khan, Y.~Fan, and M.~Chen.
\newblock Dashboard design patterns.
\newblock {\em {{IEEE} Transactions on Visualization and Computer Graphics}}, 29(1):342--352, 2023. doi: {{%
10\hspace{.1pt}\discretionary{.}{%
}{.}\hspace{.4pt}1109\discretionary{/}{%
}{/}TVCG\hspace{.1pt}\discretionary{.}{%
}{.}\hspace{.4pt}2022\hspace{.1pt}\discretionary{.}{%
}{.}\hspace{.4pt}3209448}}


\bibitem{conf/chi/BadamZSEE16}
S.~K. Badam, J.~Zhao, S.~Sen, N.~Elmqvist, and D.~S. Ebert.
\newblock {TimeFork}: Interactive prediction of time series.
\newblock In {\em Proceedings of the ACM Conference on Human Factors in Computing Systems}, pp. 5409--5420. ACM, 2016.

\bibitem{DBLP:conf/avi/BaldonadoWK00}
M.~Q.~W. Baldonado, A.~Woodruff, and A.~Kuchinsky.
\newblock Guidelines for using multiple views in information visualization.
\newblock In {\em Proceedings of the {ACM} Conference on Advanced Visual Interfaces}, pp. 110--119. {ACM}, {New York, NY, USA}, 2000. doi: {{%
10\hspace{.1pt}\discretionary{.}{%
}{.}\hspace{.4pt}1145\discretionary{/}{%
}{/}345513\hspace{.1pt}\discretionary{.}{%
}{.}\hspace{.4pt}345271}}


\bibitem{balzer05voronoitreemap}
M.~Balzer and O.~Deussen.
\newblock Voronoi treemaps.
\newblock In {\em Proceedings of the IEEE Symposium on Information Visualization}, pp. 49--56. {IEEE Computer Society}, Los Alamitos, CA, USA, 2005. doi: {{%
10\hspace{.1pt}\discretionary{.}{%
}{.}\hspace{.4pt}1109\discretionary{/}{%
}{/}INFVIS\hspace{.1pt}\discretionary{.}{%
}{.}\hspace{.4pt}2005\hspace{.1pt}\discretionary{.}{%
}{.}\hspace{.4pt}1532128}}


\bibitem{bederson02oqtreemaps}
B.~B. Bederson, B.~Shneiderman, and M.~Wattenberg.
\newblock Ordered and quantum treemaps: Making effective use of {2D} space to display hierarchies.
\newblock {\em ACM Transactions on Graphics}, 21(4):833–854, 2002. doi: {{%
10\hspace{.1pt}\discretionary{.}{%
}{.}\hspace{.4pt}1145\discretionary{/}{%
}{/}571647\hspace{.1pt}\discretionary{.}{%
}{.}\hspace{.4pt}571649}}


\bibitem{DBLP:journals/tvcg/ByronW08}
L.~Byron and M.~Wattenberg.
\newblock Stacked graphs - geometry {\&} aesthetics.
\newblock {\em {{IEEE} Transactions on Visualization and Computer Graphics}}, 14(6):1245--1252, 2008. doi: {{%
10\hspace{.1pt}\discretionary{.}{%
}{.}\hspace{.4pt}1109\discretionary{/}{%
}{/}TVCG\hspace{.1pt}\discretionary{.}{%
}{.}\hspace{.4pt}2008\hspace{.1pt}\discretionary{.}{%
}{.}\hspace{.4pt}166}}


\bibitem{cao18voila}
N.~Cao, C.~Lin, Q.~Zhu, Y.-R. Lin, X.~Teng, and X.~Wen.
\newblock Voila: Visual anomaly detection and monitoring with streaming spatiotemporal data.
\newblock {\em IEEE Transactions on Visualization and Computer Graphics}, 24(1):23--33, 2018. doi: {{%
10\hspace{.1pt}\discretionary{.}{%
}{.}\hspace{.4pt}1109\discretionary{/}{%
}{/}TVCG\hspace{.1pt}\discretionary{.}{%
}{.}\hspace{.4pt}2017\hspace{.1pt}\discretionary{.}{%
}{.}\hspace{.4pt}2744419}}


\bibitem{choi22mitovis_tvcg}
J.~Choi, H.-J. Oh, H.~Lee, S.~Kim, S.-K. Kwon, and W.-K. Jeong.
\newblock {MitoVis}: A unified visual analytics system for end-to-end neuronal mitochondria analysis.
\newblock {\em IEEE Transactions on Visualization and Computer Graphics}, pp. 1--15, 2022. doi: {{%
10\hspace{.1pt}\discretionary{.}{%
}{.}\hspace{.4pt}1109\discretionary{/}{%
}{/}TVCG\hspace{.1pt}\discretionary{.}{%
}{.}\hspace{.4pt}2022\hspace{.1pt}\discretionary{.}{%
}{.}\hspace{.4pt}3233548}}


\bibitem{choi18topicontiles}
M.~Choi, S.~Shin, J.~Choi, S.~Langevin, C.~Bethune, P.~Horne, N.~Kronenfeld, R.~Kannan, B.~Drake, H.~Park, and J.~Choo.
\newblock {TopicOnTiles}: Tile-based spatio-temporal event analytics via exclusive topic modeling on social media.
\newblock In {\em Proceedings of the {ACM} Conference on Human Factors in Computing Systems}, p. 1–11. {ACM}, {New York, NY, USA}, 2018. doi: {{%
10\hspace{.1pt}\discretionary{.}{%
}{.}\hspace{.4pt}1145\discretionary{/}{%
}{/}3173574\hspace{.1pt}\discretionary{.}{%
}{.}\hspace{.4pt}3174157}}


\bibitem{chi11textflow}
W.~Cui, S.~Liu, L.~Tan, C.~Shi, Y.~Song, Z.~Gao, H.~Qu, and X.~Tong.
\newblock {TextFlow}: Towards better understanding of evolving topics in text.
\newblock {\em {{IEEE} Transactions on Visualization and Computer Graphics}}, 17(12):2412--2421, 2011. doi: {{%
10\hspace{.1pt}\discretionary{.}{%
}{.}\hspace{.4pt}1109\discretionary{/}{%
}{/}TVCG\hspace{.1pt}\discretionary{.}{%
}{.}\hspace{.4pt}2011\hspace{.1pt}\discretionary{.}{%
}{.}\hspace{.4pt}239}}


\bibitem{doraiswamy18taxivis}
H.~Doraiswamy, J.~Freire, M.~Lage, F.~Miranda, and C.~Silva.
\newblock Spatio-temporal urban data analysis: A visual analytics perspective.
\newblock {\em IEEE Computer Graphics and Applications}, 38(5):26--35, 2018. doi: {{%
10\hspace{.1pt}\discretionary{.}{%
}{.}\hspace{.4pt}1109\discretionary{/}{%
}{/}MCG\hspace{.1pt}\discretionary{.}{%
}{.}\hspace{.4pt}2018\hspace{.1pt}\discretionary{.}{%
}{.}\hspace{.4pt}053491728}}


\bibitem{elhamdadi22affectivetda_tvcg}
H.~Elhamdadi, S.~Canavan, and P.~Rosen.
\newblock {AffectiveTDA}: Using topological data analysis to improve analysis and explainability in affective computing.
\newblock {\em {{IEEE} Transactions on Visualization and Computer Graphics}}, 28(1):769--779, 2022. doi: {{%
10\hspace{.1pt}\discretionary{.}{%
}{.}\hspace{.4pt}1109\discretionary{/}{%
}{/}TVCG\hspace{.1pt}\discretionary{.}{%
}{.}\hspace{.4pt}2021\hspace{.1pt}\discretionary{.}{%
}{.}\hspace{.4pt}3114784}}


\bibitem{Elmqvist2010}
N.~Elmqvist and J.-D. Fekete.
\newblock Hierarchical aggregation for information visualization: Overview, techniques, and design guidelines.
\newblock {\em IEEE Transactions on Visualization and Computer Graphics}, 16(3):439--454, 2010. doi: {{%
10\hspace{.1pt}\discretionary{.}{%
}{.}\hspace{.4pt}1109\discretionary{/}{%
}{/}TVCG\hspace{.1pt}\discretionary{.}{%
}{.}\hspace{.4pt}2009\hspace{.1pt}\discretionary{.}{%
}{.}\hspace{.4pt}84}}


\bibitem{DBLP:journals/tvcg/ElshehalyRBMAGR21}
M.~Elshehaly, R.~Randell, M.~Brehmer, L.~McVey, N.~Alvarado, C.~P. Gale, and R.~A. Ruddle.
\newblock {QualDash}: Adaptable generation of visualisation dashboards for healthcare quality improvement.
\newblock {\em {{IEEE} Transactions on Visualization and Computer Graphics}}, 27(2):689--699, 2021. doi: {{%
10\hspace{.1pt}\discretionary{.}{%
}{.}\hspace{.4pt}1109\discretionary{/}{%
}{/}TVCG\hspace{.1pt}\discretionary{.}{%
}{.}\hspace{.4pt}2020\hspace{.1pt}\discretionary{.}{%
}{.}\hspace{.4pt}3030424}}


\bibitem{few2006information}
S.~Few.
\newblock {\em Information Dashboard Design: The Effective Visual Communication of Data}.
\newblock O'Reilly Media, Inc., Sebastopol, CA, USA, 2006.

\bibitem{fu18tcal_chi}
S.~Fu, J.~Zhao, H.~F. Cheng, H.~Zhu, and J.~Marlow.
\newblock {T-Cal}: Understanding team conversational data with calendar-based visualization.
\newblock In {\em Proceedings of the {ACM} Conference on Human Factors in Computing Systems}, p. 1–13. {ACM}, {New York, NY, USA}, 2018. doi: {{%
10\hspace{.1pt}\discretionary{.}{%
}{.}\hspace{.4pt}1145\discretionary{/}{%
}{/}3173574\hspace{.1pt}\discretionary{.}{%
}{.}\hspace{.4pt}3174074}}


\bibitem{DBLP:journals/tvcg/GadJGEEHR15}
S.~Gad, W.~Javed, S.~Ghani, N.~Elmqvist, E.~T. Ewing, K.~N. Hampton, and N.~Ramakrishnan.
\newblock {ThemeDelta}: Dynamic segmentations over temporal topic models.
\newblock {\em {{IEEE} Transactions on Visualization and Computer Graphics}}, 21(5):672--685, 2015. doi: {{%
10\hspace{.1pt}\discretionary{.}{%
}{.}\hspace{.4pt}1109\discretionary{/}{%
}{/}TVCG\hspace{.1pt}\discretionary{.}{%
}{.}\hspace{.4pt}2014\hspace{.1pt}\discretionary{.}{%
}{.}\hspace{.4pt}2388208}}


\bibitem{garciazanabria22cripav_tvcg}
G.~García-Zanabria, M.~M. Raimundo, J.~Poco, M.~B. Nery, C.~T. Silva, S.~Adorno, and L.~G. Nonato.
\newblock {CriPAV}: Street-level crime patterns analysis and visualization.
\newblock {\em {{IEEE} Transactions on Visualization and Computer Graphics}}, 28(12):4000--4015, 2022. doi: {{%
10\hspace{.1pt}\discretionary{.}{%
}{.}\hspace{.4pt}1109\discretionary{/}{%
}{/}TVCG\hspace{.1pt}\discretionary{.}{%
}{.}\hspace{.4pt}2021\hspace{.1pt}\discretionary{.}{%
}{.}\hspace{.4pt}3111146}}


\bibitem{gotz20vahieragg}
D.~Gotz, J.~Zhang, W.~Wang, J.~Shrestha, and D.~Borland.
\newblock Visual analysis of high-dimensional event sequence data via dynamic hierarchical aggregation.
\newblock {\em IEEE Transactions on Visualization and Computer Graphics}, 26(1):440--450, 2020. doi: {{%
10\hspace{.1pt}\discretionary{.}{%
}{.}\hspace{.4pt}1109\discretionary{/}{%
}{/}TVCG\hspace{.1pt}\discretionary{.}{%
}{.}\hspace{.4pt}2019\hspace{.1pt}\discretionary{.}{%
}{.}\hspace{.4pt}2934661}}


\bibitem{Greenberg1985}
S.~Greenberg and I.~H. Witten.
\newblock Adaptive personalized interfaces—a question of viability.
\newblock {\em Behaviour \& Information Technology}, 4(1):31--45, 1985. doi: {{%
10\hspace{.1pt}\discretionary{.}{%
}{.}\hspace{.4pt}1080\discretionary{/}{%
}{/}01449298508901785}}


\bibitem{havre02themeriver}
S.~Havre, E.~Hetzler, P.~Whitney, and L.~Nowell.
\newblock {ThemeRiver}: visualizing thematic changes in large document collections.
\newblock {\em {{IEEE} Transactions on Visualization and Computer Graphics}}, 8(1):9--20, 2002. doi: {{%
10\hspace{.1pt}\discretionary{.}{%
}{.}\hspace{.4pt}1109\discretionary{/}{%
}{/}2945\hspace{.1pt}\discretionary{.}{%
}{.}\hspace{.4pt}981848}}


\bibitem{heimerl16citerivers}
F.~Heimerl, Q.~Han, S.~Koch, and T.~Ertl.
\newblock {CiteRivers}: Visual analytics of citation patterns.
\newblock {\em {{IEEE} Transactions on Visualization and Computer Graphics}}, 22(1):190--199, 2016. doi: {{%
10\hspace{.1pt}\discretionary{.}{%
}{.}\hspace{.4pt}1109\discretionary{/}{%
}{/}TVCG\hspace{.1pt}\discretionary{.}{%
}{.}\hspace{.4pt}2015\hspace{.1pt}\discretionary{.}{%
}{.}\hspace{.4pt}2467621}}


\bibitem{henry07nodetrix}
N.~Henry, J.-D. Fekete, and M.~J. McGuffin.
\newblock Nodetrix: a hybrid visualization of social networks.
\newblock {\em IEEE Transactions on Visualization and Computer Graphics}, 13(6):1302--1309, 2007. doi: {{%
10\hspace{.1pt}\discretionary{.}{%
}{.}\hspace{.4pt}1109\discretionary{/}{%
}{/}TVCG\hspace{.1pt}\discretionary{.}{%
}{.}\hspace{.4pt}2007\hspace{.1pt}\discretionary{.}{%
}{.}\hspace{.4pt}70582}}


\bibitem{hohman19gamut}
F.~Hohman, A.~Head, R.~Caruana, R.~DeLine, and S.~M. Drucker.
\newblock Gamut: A design probe to understand how data scientists understand machine learning models.
\newblock In {\em Proceedings of the {ACM} Conference on Human Factors in Computing Systems}, p. 579:1–579:13. {ACM}, {New York, NY, USA}, 2019. doi: {{%
10\hspace{.1pt}\discretionary{.}{%
}{.}\hspace{.4pt}1145\discretionary{/}{%
}{/}3290605\hspace{.1pt}\discretionary{.}{%
}{.}\hspace{.4pt}3300809}}


\bibitem{hoque23visualconcept_tvcg}
M.~N. Hoque, W.~He, A.~K. Shekar, L.~Gou, and L.~Ren.
\newblock Visual concept programming: A visual analytics approach to injecting human intelligence at scale.
\newblock {\em IEEE Transactions on Visualization and Computer Graphics}, 29(1):74--83, 2023. doi: {{%
10\hspace{.1pt}\discretionary{.}{%
}{.}\hspace{.4pt}1109\discretionary{/}{%
}{/}TVCG\hspace{.1pt}\discretionary{.}{%
}{.}\hspace{.4pt}2022\hspace{.1pt}\discretionary{.}{%
}{.}\hspace{.4pt}3209466}}


\bibitem{huang23conceptexplainer_tvcg}
J.~Huang, A.~Mishra, B.~C. Kwon, and C.~Bryan.
\newblock {ConceptExplainer}: Interactive explanation for deep neural networks from a concept perspective.
\newblock {\em {{IEEE} Transactions on Visualization and Computer Graphics}}, 29(1):831--841, 2023. doi: {{%
10\hspace{.1pt}\discretionary{.}{%
}{.}\hspace{.4pt}1109\discretionary{/}{%
}{/}TVCG\hspace{.1pt}\discretionary{.}{%
}{.}\hspace{.4pt}2022\hspace{.1pt}\discretionary{.}{%
}{.}\hspace{.4pt}3209384}}


\bibitem{DBLP:conf/apvis/JavedE12}
W.~Javed and N.~Elmqvist.
\newblock Exploring the design space of composite visualization.
\newblock In {\em Proceedings of the {IEEE} Pacific Symposium on Visualization}, pp. 1--8. {IEEE Computer Society}, Los Alamitos, CA, USA, 2012. doi: {{%
10\hspace{.1pt}\discretionary{.}{%
}{.}\hspace{.4pt}1109\discretionary{/}{%
}{/}PACIFICVIS\hspace{.1pt}\discretionary{.}{%
}{.}\hspace{.4pt}2012\hspace{.1pt}\discretionary{.}{%
}{.}\hspace{.4pt}6183556}}


\bibitem{Javed2013}
W.~Javed and N.~Elmqvist.
\newblock {ExPlates}: Spatializing interactive analysis to scaffold visual exploration.
\newblock {\em Computer Graphics Forum}, 32(2):441--450, 2013. doi: {{%
10\hspace{.1pt}\discretionary{.}{%
}{.}\hspace{.4pt}1111\discretionary{/}{%
}{/}cgf\hspace{.1pt}\discretionary{.}{%
}{.}\hspace{.4pt}12131}}


\bibitem{Javed2012a}
W.~Javed, S.~Ghani, and N.~Elmqvist.
\newblock {GravNav}: Using a gravity model for multi-scale navigation.
\newblock In {\em Proceedings of the {ACM} Conference on Advanced Visual Interfaces}, pp. 217--224. {ACM}, {New York, NY, USA}, 2012. doi: {{%
10\hspace{.1pt}\discretionary{.}{%
}{.}\hspace{.4pt}1145\discretionary{/}{%
}{/}2254556\hspace{.1pt}\discretionary{.}{%
}{.}\hspace{.4pt}2254597}}


\bibitem{jin23trafficvis_tvcg}
S.~Jin, H.~Lee, C.~Park, H.~Chu, Y.~Tae, J.~Choo, and S.~Ko.
\newblock A visual analytics system for improving attention-based traffic forecasting models.
\newblock {\em IEEE Transactions on Visualization and Computer Graphics}, 29(1):1102--1112, 2023. doi: {{%
10\hspace{.1pt}\discretionary{.}{%
}{.}\hspace{.4pt}1109\discretionary{/}{%
}{/}TVCG\hspace{.1pt}\discretionary{.}{%
}{.}\hspace{.4pt}2022\hspace{.1pt}\discretionary{.}{%
}{.}\hspace{.4pt}3209462}}


\bibitem{johnson91treemaps}
B.~Johnson and B.~Shneiderman.
\newblock Tree-maps: A space-filling approach to the visualization of hierarchical information structures.
\newblock In {\em Proceedings of the IEEE Conference on Visualization}, p. 284–291. {IEEE Computer Society}, Los Alamitos, CA, USA, 1991.

\bibitem{kim17topiclens}
M.~Kim, K.~Kang, D.~Park, J.~Choo, and N.~Elmqvist.
\newblock {TopicLens}: Efficient multi-level visual topic exploration of large-scale document collections.
\newblock {\em {{IEEE} Transactions on Visualization and Computer Graphics}}, 23(1):151--160, 2017. doi: {{%
10\hspace{.1pt}\discretionary{.}{%
}{.}\hspace{.4pt}1109\discretionary{/}{%
}{/}TVCG\hspace{.1pt}\discretionary{.}{%
}{.}\hspace{.4pt}2016\hspace{.1pt}\discretionary{.}{%
}{.}\hspace{.4pt}2598445}}


\bibitem{DBLP:conf/chi/LampingRP95}
J.~Lamping, R.~Rao, and P.~Pirolli.
\newblock A focus+context technique based on hyperbolic geometry for visualizing large hierarchies.
\newblock In {\em Proceedings of the {ACM} Conference on Human Factors in Computing Systems}, pp. 401--408. {ACM}, {New York, NY, USA}, 1995. doi: {{%
10\hspace{.1pt}\discretionary{.}{%
}{.}\hspace{.4pt}1145\discretionary{/}{%
}{/}223904\hspace{.1pt}\discretionary{.}{%
}{.}\hspace{.4pt}223956}}


\bibitem{li20maravis_chi}
Q.~Li, H.~Lin, X.~Wei, Y.~Huang, L.~Fan, J.~Du, X.~Ma, and T.~Chen.
\newblock {MaraVis}: Representation and coordinated intervention of medical encounters in urban marathon.
\newblock In {\em Proceedings of the {ACM} Conference on Human Factors in Computing Systems}, p. 1–12. {ACM}, {New York, NY, USA}, 2020. doi: {{%
10\hspace{.1pt}\discretionary{.}{%
}{.}\hspace{.4pt}1145\discretionary{/}{%
}{/}3313831\hspace{.1pt}\discretionary{.}{%
}{.}\hspace{.4pt}3376281}}


\bibitem{lins13nanocubes}
L.~Lins, J.~T. Klosowski, and C.~Scheidegger.
\newblock Nanocubes for real-time exploration of spatiotemporal datasets.
\newblock {\em IEEE Transactions on Visualization and Computer Graphics}, 19(12):2456--2465, 2013. doi: {{%
10\hspace{.1pt}\discretionary{.}{%
}{.}\hspace{.4pt}1109\discretionary{/}{%
}{/}TVCG\hspace{.1pt}\discretionary{.}{%
}{.}\hspace{.4pt}2013\hspace{.1pt}\discretionary{.}{%
}{.}\hspace{.4pt}179}}


\bibitem{DBLP:journals/tvcg/MaMGHZXDWC21}
R.~Ma, H.~Mei, H.~Guan, W.~Huang, F.~Zhang, C.~Xin, W.~Dai, X.~Wen, and W.~Chen.
\newblock {LADV:} deep learning assisted authoring of dashboard visualizations from images and sketches.
\newblock {\em {{IEEE} Transactions on Visualization and Computer Graphics}}, 27(9):3717--3732, 2021. doi: {{%
10\hspace{.1pt}\discretionary{.}{%
}{.}\hspace{.4pt}1109\discretionary{/}{%
}{/}TVCG\hspace{.1pt}\discretionary{.}{%
}{.}\hspace{.4pt}2020\hspace{.1pt}\discretionary{.}{%
}{.}\hspace{.4pt}2980227}}


\bibitem{DBLP:journals/cgf/MacNeilE13}
S.~MacNeil and N.~Elmqvist.
\newblock Visualization mosaics for multivariate visual exploration.
\newblock {\em Computer Graphics Forum}, 32(6):38--50, 2013. doi: {{%
10\hspace{.1pt}\discretionary{.}{%
}{.}\hspace{.4pt}1111\discretionary{/}{%
}{/}CGF\hspace{.1pt}\discretionary{.}{%
}{.}\hspace{.4pt}12013}}


\bibitem{marcus11twitinfo}
A.~Marcus, M.~S. Bernstein, O.~Badar, D.~R. Karger, S.~Madden, and R.~C. Miller.
\newblock Twitinfo: Aggregating and visualizing microblogs for event exploration.
\newblock In {\em Proceedings of the {ACM} Conference on Human Factors in Computing Systems}, p. 227–236. {ACM}, {New York, NY, USA}, 2011. doi: {{%
10\hspace{.1pt}\discretionary{.}{%
}{.}\hspace{.4pt}1145\discretionary{/}{%
}{/}1978942\hspace{.1pt}\discretionary{.}{%
}{.}\hspace{.4pt}1978975}}


\bibitem{DBLP:journals/cgf/MathisenHKGE19}
A.~Mathisen, T.~Horak, C.~N. Klokmose, K.~Gr{\o}nb{\ae}k, and N.~Elmqvist.
\newblock {InsideInsights}: Integrating data-driven reporting in collaborative visual analytics.
\newblock {\em Computer Graphics Forum}, 38(3):649--661, 2019. doi: {{%
10\hspace{.1pt}\discretionary{.}{%
}{.}\hspace{.4pt}1111\discretionary{/}{%
}{/}CGF\hspace{.1pt}\discretionary{.}{%
}{.}\hspace{.4pt}13717}}


\bibitem{DBLP:journals/tvcg/McDonnelE09}
B.~McDonnel and N.~Elmqvist.
\newblock Towards utilizing {GPUs} in information visualization: {A} model and implementation of image-space operations.
\newblock {\em {{IEEE} Transactions on Visualization and Computer Graphics}}, 15(6):1105--1112, 2009. doi: {{%
10\hspace{.1pt}\discretionary{.}{%
}{.}\hspace{.4pt}1109\discretionary{/}{%
}{/}TVCG\hspace{.1pt}\discretionary{.}{%
}{.}\hspace{.4pt}2009\hspace{.1pt}\discretionary{.}{%
}{.}\hspace{.4pt}191}}


\bibitem{mcneill23zombievis_chi}
G.~Mcneill, M.~Sondag, S.~Powell, P.~Asplin, C.~Turkay, F.~Moller, and D.~Archambault.
\newblock From asymptomatics to zombies: Visualization-based education of disease modeling for children.
\newblock In {\em Proceedings of the {ACM} Conference on Human Factors in Computing Systems}. {ACM}, {New York, NY, USA}, 2023. doi: {{%
10\hspace{.1pt}\discretionary{.}{%
}{.}\hspace{.4pt}1145\discretionary{/}{%
}{/}3544548\hspace{.1pt}\discretionary{.}{%
}{.}\hspace{.4pt}3581573}}


\bibitem{mendez21academic_chi}
G.~Mendez, L.~Gal\'{a}rraga, and K.~Chiluiza.
\newblock Showing academic performance predictions during term planning: Effects on students’ decisions, behaviors, and preferences.
\newblock In {\em Proceedings of the {ACM} Conference on Human Factors in Computing Systems}. {ACM}, {New York, NY, USA}, 2021. doi: {{%
10\hspace{.1pt}\discretionary{.}{%
}{.}\hspace{.4pt}1145\discretionary{/}{%
}{/}3411764\hspace{.1pt}\discretionary{.}{%
}{.}\hspace{.4pt}3445718}}


\bibitem{noonpakdee18dashboardbi}
W.~Noonpakdee, T.~Khunkornsiri, A.~Phothichai, and K.~Danaisawat.
\newblock A framework for analyzing and developing dashboard templates for small and medium enterprises.
\newblock In {\em Proceedings of the IEEE Conference on Industrial Engineering and Applications}, pp. 479--483. {IEEE}, Piscataway, NJ, USA, 2018. doi: {{%
10\hspace{.1pt}\discretionary{.}{%
}{.}\hspace{.4pt}1109\discretionary{/}{%
}{/}IEA\hspace{.1pt}\discretionary{.}{%
}{.}\hspace{.4pt}2018\hspace{.1pt}\discretionary{.}{%
}{.}\hspace{.4pt}8387148}}


\bibitem{pahins17hashedcubes}
C.~A.~L. Pahins, S.~A. Stephens, C.~Scheidegger, and J.~L.~D. Comba.
\newblock Hashedcubes: Simple, low memory, real-time visual exploration of big data.
\newblock {\em {{IEEE} Transactions on Visualization and Computer Graphics}}, 23(1):671--680, 2017. doi: {{%
10\hspace{.1pt}\discretionary{.}{%
}{.}\hspace{.4pt}1109\discretionary{/}{%
}{/}TVCG\hspace{.1pt}\discretionary{.}{%
}{.}\hspace{.4pt}2016\hspace{.1pt}\discretionary{.}{%
}{.}\hspace{.4pt}2598624}}


\bibitem{DBLP:journals/tvcg/PandeySS23}
A.~Pandey, A.~Srinivasan, and V.~Setlur.
\newblock {MEDLEY:} intent-based recommendations to support dashboard composition.
\newblock {\em {{IEEE} Transactions on Visualization and Computer Graphics}}, 29(1):1135--1145, 2023. doi: {{%
10\hspace{.1pt}\discretionary{.}{%
}{.}\hspace{.4pt}1109\discretionary{/}{%
}{/}TVCG\hspace{.1pt}\discretionary{.}{%
}{.}\hspace{.4pt}2022\hspace{.1pt}\discretionary{.}{%
}{.}\hspace{.4pt}3209421}}


\bibitem{DBLP:journals/ivs/ParkSZDRE22}
D.~G. Park, M.~Suhail, M.~Zheng, C.~Dunne, E.~D. Ragan, and N.~Elmqvist.
\newblock {StoryFacets}: {A} design study on storytelling with visualizations for collaborative data analysis.
\newblock {\em Information Visualization}, 21(1):3--16, 2022. doi: {{%
10\hspace{.1pt}\discretionary{.}{%
}{.}\hspace{.4pt}1177\discretionary{/}{%
}{/}14738716211032653}}


\bibitem{piringer10hierscatterplot}
H.~Piringer, M.~Buchetics, H.~Hauser, and E.~Gr\"{o}ller.
\newblock Hierarchical difference scatterplots: Interactive visual analysis of data cubes.
\newblock {\em SIGKDD Explorations Newsletter}, 11(2):49–58, 2010. doi: {{%
10\hspace{.1pt}\discretionary{.}{%
}{.}\hspace{.4pt}1145\discretionary{/}{%
}{/}1809400\hspace{.1pt}\discretionary{.}{%
}{.}\hspace{.4pt}1809408}}


\bibitem{DBLP:conf/infovis/PlaisantGB02}
C.~Plaisant, J.~Grosjean, and B.~B. Bederson.
\newblock {SpaceTree}: Supporting exploration in large node link tree, design evolution and empirical evaluation.
\newblock In {\em Proceedings of the {IEEE} Symposium on Information Visualization}, pp. 57--64. {IEEE Computer Society}, Los Alamitos, CA, USA, 2002. doi: {{%
10\hspace{.1pt}\discretionary{.}{%
}{.}\hspace{.4pt}1109\discretionary{/}{%
}{/}INFVIS\hspace{.1pt}\discretionary{.}{%
}{.}\hspace{.4pt}2002\hspace{.1pt}\discretionary{.}{%
}{.}\hspace{.4pt}1173148}}


\bibitem{DBLP:journals/tvcg/QuH18}
Z.~Qu and J.~Hullman.
\newblock Keeping multiple views consistent: Constraints, validations, and exceptions in visualization authoring.
\newblock {\em {{IEEE} Transactions on Visualization and Computer Graphics}}, 24(1):468--477, 2018. doi: {{%
10\hspace{.1pt}\discretionary{.}{%
}{.}\hspace{.4pt}1109\discretionary{/}{%
}{/}TVCG\hspace{.1pt}\discretionary{.}{%
}{.}\hspace{.4pt}2017\hspace{.1pt}\discretionary{.}{%
}{.}\hspace{.4pt}2744198}}


\bibitem{reuter15xhelp_chi}
C.~Reuter, T.~Ludwig, M.-A. Kaufhold, and V.~Pipek.
\newblock {XHELP}: Design of a cross-platform social-media application to support volunteer moderators in disasters.
\newblock In {\em Proceedings of the {ACM} Conference on Human Factors in Computing Systems}, p. 4093–4102. {ACM}, {New York, NY, USA}, 2015. doi: {{%
10\hspace{.1pt}\discretionary{.}{%
}{.}\hspace{.4pt}1145\discretionary{/}{%
}{/}2702123\hspace{.1pt}\discretionary{.}{%
}{.}\hspace{.4pt}2702171}}


\bibitem{Roberts2007}
J.~C. Roberts.
\newblock State of the art: Coordinated \& multiple views in exploratory visualization.
\newblock In {\em Proceedings of the International Conference on Coordinated and Multiple Views in Exploratory Visualization}, pp. 61--71. {IEEE}, Piscataway, NJ, USA, 2007. doi: {{%
10\hspace{.1pt}\discretionary{.}{%
}{.}\hspace{.4pt}1109\discretionary{/}{%
}{/}cmv\hspace{.1pt}\discretionary{.}{%
}{.}\hspace{.4pt}2007\hspace{.1pt}\discretionary{.}{%
}{.}\hspace{.4pt}20}}


\bibitem{samrose21meetingcoach_chi}
S.~Samrose, D.~McDuff, R.~Sim, J.~Suh, K.~Rowan, J.~Hernandez, S.~Rintel, K.~Moynihan, and M.~Czerwinski.
\newblock {MeetingCoach}: An intelligent dashboard for supporting effective and inclusive meetings.
\newblock In {\em Proceedings of the {ACM} Conference on Human Factors in Computing Systems}. {ACM}, {New York, NY, USA}, 2021. doi: {{%
10\hspace{.1pt}\discretionary{.}{%
}{.}\hspace{.4pt}1145\discretionary{/}{%
}{/}3411764\hspace{.1pt}\discretionary{.}{%
}{.}\hspace{.4pt}3445615}}


\bibitem{sarikaya19dashboards_tvcg}
A.~Sarikaya, M.~Correll, L.~Bartram, M.~Tory, and D.~Fisher.
\newblock What do we talk about when we talk about dashboards?
\newblock {\em IEEE Transactions on Visualization and Computer Graphics}, 25(1):682--692, 2019. doi: {{%
10\hspace{.1pt}\discretionary{.}{%
}{.}\hspace{.4pt}1109\discretionary{/}{%
}{/}TVCG\hspace{.1pt}\discretionary{.}{%
}{.}\hspace{.4pt}2018\hspace{.1pt}\discretionary{.}{%
}{.}\hspace{.4pt}2864903}}


\bibitem{AUIbook1993}
M.~Schneider-Hufschmidt, T.~Kühme, and U.~Malinowski, eds.
\newblock {\em Adaptive User Interfaces: Principles and Practice}.
\newblock Elsevier Science, Amsterdam, North-Holland, 1993.

\bibitem{shneiderman92treemap}
B.~Shneiderman.
\newblock Tree visualization with tree-maps: {2-D} space-filling approach.
\newblock {\em ACM Transactions on Graphics}, 11(1):92–99, Jan. 1992. doi: {{%
10\hspace{.1pt}\discretionary{.}{%
}{.}\hspace{.4pt}1145\discretionary{/}{%
}{/}102377\hspace{.1pt}\discretionary{.}{%
}{.}\hspace{.4pt}115768}}


\bibitem{Srinivasan2023}
A.~Srinivasan and V.~Setlur.
\newblock {BOLT}: {A} natural language interface for dashboard authoring.
\newblock In {\em Short Paper Proceedings of the IEEE VGTC/Eurographics Symposium on Visualization}, pp. 7--11, 2023. doi: {{%
10\hspace{.1pt}\discretionary{.}{%
}{.}\hspace{.4pt}2312\discretionary{/}{%
}{/}evs20231035}}


\bibitem{stolte02polaris_tvcg}
C.~Stolte, D.~Tang, and P.~Hanrahan.
\newblock Polaris: a system for query, analysis, and visualization of multidimensional relational databases.
\newblock {\em IEEE Transactions on Visualization and Computer Graphics}, 8(1):52--65, 2002. doi: {{%
10\hspace{.1pt}\discretionary{.}{%
}{.}\hspace{.4pt}1109\discretionary{/}{%
}{/}2945\hspace{.1pt}\discretionary{.}{%
}{.}\hspace{.4pt}981851}}


\bibitem{Stolte2003}
C.~Stolte, D.~Tang, and P.~Hanrahan.
\newblock Multiscale visualization using data cubes.
\newblock {\em {{IEEE} Transactions on Visualization and Computer Graphics}}, 9(2):176--187, 2003.

\bibitem{thanyadit23tutorvis_chi}
S.~Thanyadit, M.~Heintz, and E.~L.-C. Law.
\newblock Tutor in-sight: Guiding and visualizing students’ attention with mixed reality avatar presentation tools.
\newblock In {\em Proceedings of the {ACM} Conference on Human Factors in Computing Systems}. {ACM}, {New York, NY, USA}, 2023. doi: {{%
10\hspace{.1pt}\discretionary{.}{%
}{.}\hspace{.4pt}1145\discretionary{/}{%
}{/}3544548\hspace{.1pt}\discretionary{.}{%
}{.}\hspace{.4pt}3581069}}


\bibitem{vankollenburg18healthprof_chi}
J.~van Kollenburg, S.~Bogers, H.~Rutjes, E.~Deckers, J.~Frens, and C.~Hummels.
\newblock Exploring the value of parent tracked baby data in interactions with healthcare professionals: A data-enabled design exploration.
\newblock In {\em Proceedings of the {ACM} Conference on Human Factors in Computing Systems}, p. 1–12. {ACM}, {New York, NY, USA}, 2018. doi: {{%
10\hspace{.1pt}\discretionary{.}{%
}{.}\hspace{.4pt}1145\discretionary{/}{%
}{/}3173574\hspace{.1pt}\discretionary{.}{%
}{.}\hspace{.4pt}3173871}}


\bibitem{Viegas2006}
F.~B. Vi{\'e}gas and M.~Wattenberg.
\newblock Communication-minded visualization: {A} call to action.
\newblock {\em IBM Systems Journal}, 45(4):801--812, Apr. 2006.

\bibitem{wang22vaod_tvcg}
J.~Wang, Y.~Li, Z.~Zhou, C.~Wang, Y.~Hou, L.~Zhang, X.~Xue, M.~Kamp, X.~Zhang, and S.~Chen.
\newblock When, where and how does it fail? {A} spatial-temporal visual analytics approach for interpretable object detection in autonomous driving.
\newblock {\em {{IEEE} Transactions on Visualization and Computer Graphics}}, pp. 1--16, 2022. doi: {{%
10\hspace{.1pt}\discretionary{.}{%
}{.}\hspace{.4pt}1109\discretionary{/}{%
}{/}TVCG\hspace{.1pt}\discretionary{.}{%
}{.}\hspace{.4pt}2022\hspace{.1pt}\discretionary{.}{%
}{.}\hspace{.4pt}3201101}}


\bibitem{wang21neuralcubes}
Z.~Wang, D.~Cashman, M.~Li, J.~Li, M.~Berger, J.~A. Levine, R.~Chang, and C.~Scheidegger.
\newblock {NeuralCubes}: Deep representations for visual data exploration.
\newblock In {\em Proceedings of the International Conference on Big Data}, pp. 550--561. {IEEE Computer Society}, Los Alamitos, CA, USA, 2021. doi: {{%
10\hspace{.1pt}\discretionary{.}{%
}{.}\hspace{.4pt}1109\discretionary{/}{%
}{/}BigData52589\hspace{.1pt}\discretionary{.}{%
}{.}\hspace{.4pt}2021\hspace{.1pt}\discretionary{.}{%
}{.}\hspace{.4pt}9671390}}


\bibitem{wei10tiara}
F.~Wei, S.~Liu, Y.~Song, S.~Pan, M.~X. Zhou, W.~Qian, L.~Shi, L.~Tan, and Q.~Zhang.
\newblock {TIARA}: A visual exploratory text analytic system.
\newblock In {\em Proceedings of the ACM Conference on Knowledge Discovery and Data Mining}, p. 153–162. {ACM}, {New York, NY, USA}, 2010. doi: {{%
10\hspace{.1pt}\discretionary{.}{%
}{.}\hspace{.4pt}1145\discretionary{/}{%
}{/}1835804\hspace{.1pt}\discretionary{.}{%
}{.}\hspace{.4pt}1835827}}


\bibitem{DBLP:journals/tvcg/WuWZHZQZ22}
A.~Wu, Y.~Wang, M.~Zhou, X.~He, H.~Zhang, H.~Qu, and D.~Zhang.
\newblock {MultiVision}: {D}esigning analytical dashboards with deep learning based recommendation.
\newblock {\em {{IEEE} Transactions on Visualization and Computer Graphics}}, 28(1):162--172, 2022. doi: {{%
10\hspace{.1pt}\discretionary{.}{%
}{.}\hspace{.4pt}1109\discretionary{/}{%
}{/}TVCG\hspace{.1pt}\discretionary{.}{%
}{.}\hspace{.4pt}2021\hspace{.1pt}\discretionary{.}{%
}{.}\hspace{.4pt}3114826}}


\bibitem{Yalcin2017b}
M.~A. Yal{\c c}in, N.~Elmqvist, and B.~B. Bederson.
\newblock Keshif: Rapid and expressive tabular data exploration for novices.
\newblock {\em IEEE Transactions on Visualization and Computer Graphics}, 24(8):2339--2352, aug 2018. doi: {{%
10\hspace{.1pt}\discretionary{.}{%
}{.}\hspace{.4pt}1109\discretionary{/}{%
}{/}TVCG\hspace{.1pt}\discretionary{.}{%
}{.}\hspace{.4pt}2017\hspace{.1pt}\discretionary{.}{%
}{.}\hspace{.4pt}2723393}}


\bibitem{yan20silva_chi}
J.~N. Yan, Z.~Gu, H.~Lin, and J.~M. Rzeszotarski.
\newblock Silva: Interactively assessing machine learning fairness using causality.
\newblock In {\em Proceedings of the {ACM} Conference on Human Factors in Computing Systems}, p. 1–13. {ACM}, {New York, NY, USA}, 2020. doi: {{%
10\hspace{.1pt}\discretionary{.}{%
}{.}\hspace{.4pt}1145\discretionary{/}{%
}{/}3313831\hspace{.1pt}\discretionary{.}{%
}{.}\hspace{.4pt}3376447}}


\bibitem{yang23epimob_tvcg}
C.~Yang, Z.~Zhang, Z.~Fan, R.~Jiang, Q.~Chen, X.~Song, and R.~Shibasaki.
\newblock {EpiMob}: Interactive visual analytics of citywide human mobility restrictions for epidemic control.
\newblock {\em {{IEEE} Transactions on Visualization and Computer Graphics}}, 29(8):3586--3601, 2023. doi: {{%
10\hspace{.1pt}\discretionary{.}{%
}{.}\hspace{.4pt}1109\discretionary{/}{%
}{/}TVCG\hspace{.1pt}\discretionary{.}{%
}{.}\hspace{.4pt}2022\hspace{.1pt}\discretionary{.}{%
}{.}\hspace{.4pt}3165385}}


\bibitem{yang02interring}
J.~Yang, M.~Ward, and E.~Rundensteiner.
\newblock {InterRing}: an interactive tool for visually navigating and manipulating hierarchical structures.
\newblock In {\em Proceedings of the IEEE Symposium on Information Visualization}, pp. 77--84. {IEEE Computer Society}, Los Alamitos, CA, USA, 2002. doi: {{%
10\hspace{.1pt}\discretionary{.}{%
}{.}\hspace{.4pt}1109\discretionary{/}{%
}{/}INFVIS\hspace{.1pt}\discretionary{.}{%
}{.}\hspace{.4pt}2002\hspace{.1pt}\discretionary{.}{%
}{.}\hspace{.4pt}1173151}}


\bibitem{zhang21mapcovid}
Y.~Zhang, Y.~Sun, L.~Padilla, S.~Barua, E.~Bertini, and A.~G. Parker.
\newblock Mapping the landscape of {COVID-19} crisis visualizations.
\newblock In {\em Proceedings of the {ACM} Conference on Human Factors in Computing Systems}. {ACM}, {New York, NY, USA}, 2021. doi: {{%
10\hspace{.1pt}\discretionary{.}{%
}{.}\hspace{.4pt}1145\discretionary{/}{%
}{/}3411764\hspace{.1pt}\discretionary{.}{%
}{.}\hspace{.4pt}3445381}}


\end{thebibliography}

\begin{IEEEbiography}[{\includegraphics[width=1in,height=1.25in,clip,keepaspectratio]{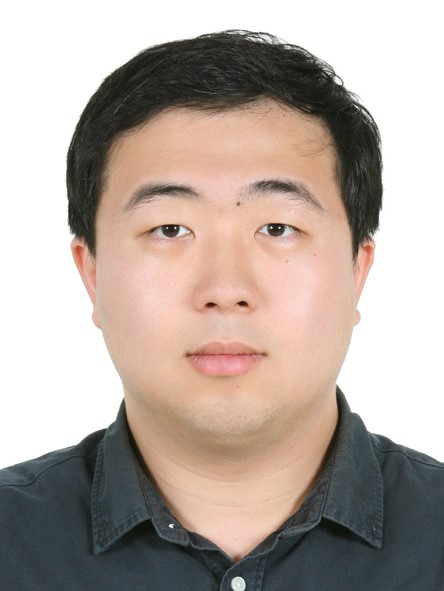}}]{Sungbok Shin}
received the Ph.D. degree in Computer Science from the University of Maryland, College Park, MD, USA, in 2024.
He is a postdoctoral researcher at Inria Saclay Centre in Saclay, France.
His research interest is in human-centered AI, visualization, HCI, and AI applications. \end{IEEEbiography}

\begin{IEEEbiography}[{\includegraphics[width=1in,height=1.25in,clip,keepaspectratio]{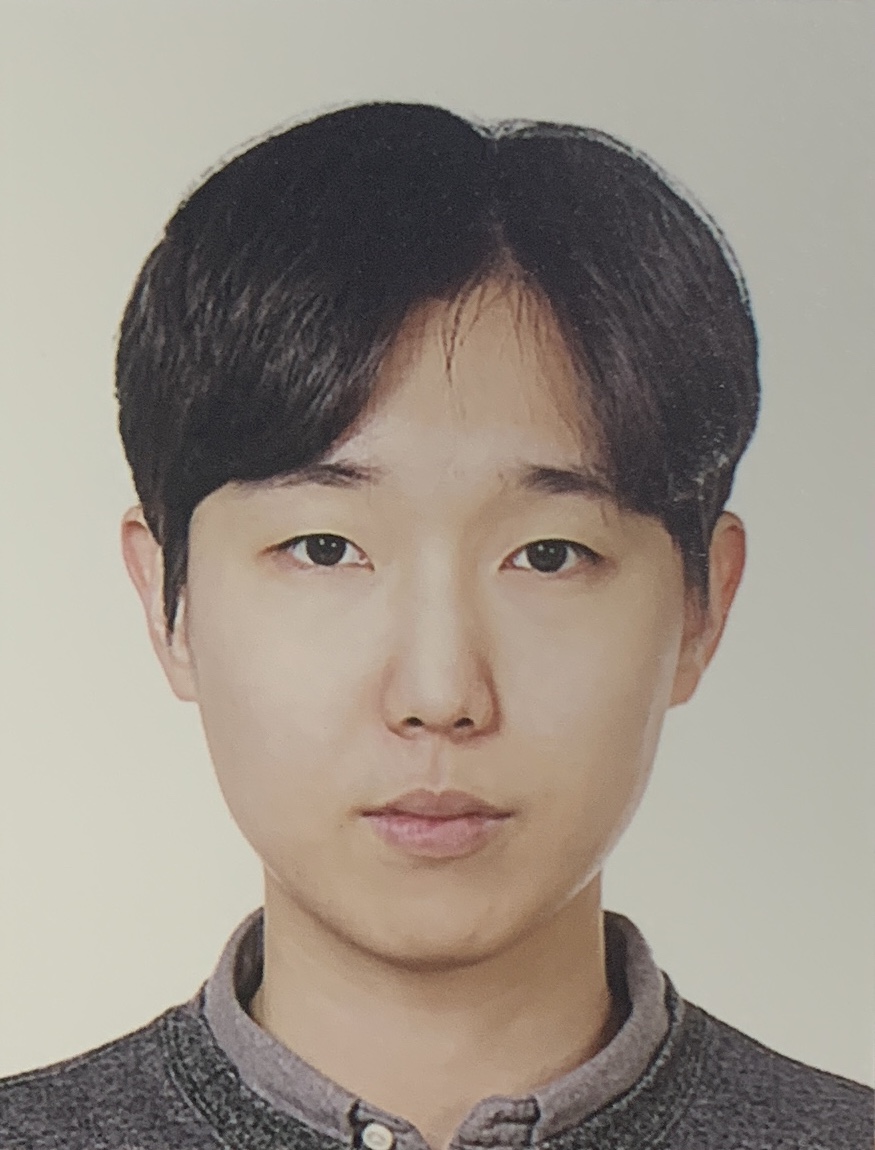}}]{Inyoup Na}
received the B.S. degree in Computer Science and Engineering from Korea University, Seoul, South Korea.
He is currently a systems engineer at an IT company, and an independent researcher in the field of HCI and visualization. \end{IEEEbiography}

\begin{IEEEbiography}[{\includegraphics[width=1in,height=1.25in,clip,keepaspectratio]{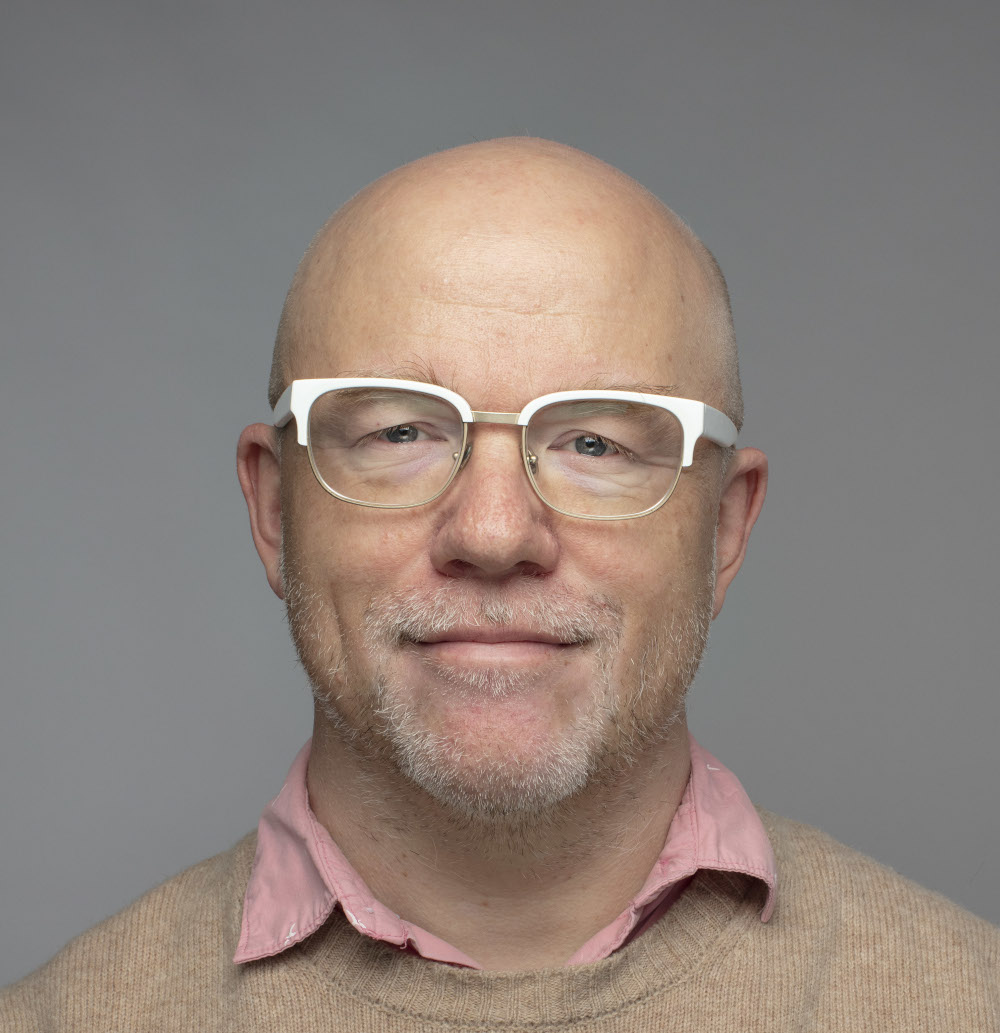}}]{Niklas Elmqvist}
received the Ph.D.\ degree in 2006 from Chalmers University of Technology in G\"{o}teborg, Sweden.
He is a Villum Investigator and professor in the Department of Computer Science at Aarhus University in Aarhus, Denmark.
He was previously faculty at University of Maryland, College Park from 2014 to 2023, and at Purdue University from 2008 to 2014. 
His research interests include visualization, HCI, and human-centered AI.
He is a Fellow of the IEEE Computer Society and the ACM.\end{IEEEbiography}

\vfill

\end{document}